\documentclass[conference,10pt]{IEEEtran}
\usepackage{epsfig}
\usepackage{times}
\usepackage{float}
\usepackage{afterpage}
\usepackage{amsmath}
\usepackage{amstext}
\usepackage{amssymb,bm}
\usepackage{latexsym}
\usepackage{color}
\usepackage{graphicx, caption, subcaption}
\usepackage{amsmath}
\usepackage{amsthm}
\usepackage{pstricks}
\usepackage{booktabs}
\usepackage{multicol}
\usepackage{lipsum}
\usepackage{dblfloatfix}
\usepackage{tikz}
\usepackage{pgfplots}

\allowdisplaybreaks

\def \eu{\mathrm{e}}

\newcommand{\mmse}{\mathrm{mmse}}

\newcommand{\E}{\mathbb{E}}

\newcommand{\X}{{\bf X}}

\newcommand{\Y}{{\bf Y}}

\newtheorem{thm}{Theorem}
\newtheorem{cor}[thm]{Corollary}
\newtheorem{lem}[thm]{Lemma}

\newtheorem{rem}{Remark}
\newtheorem{defin}{Definition}

\usepackage{enumitem}

\IEEEoverridecommandlockouts   

\begin{document}

\title{Estimating Noisy Order Statistics}

\author{
\IEEEauthorblockN{Alex Dytso$^{\dagger}$, Martina Cardone$^*$, and H. Vincent Poor$^{\dagger}$}
$^{\dagger}$ Princeton University, Princeton, NJ 08544, USA,
Email: \{adytso, poor\}@princeton.edu\\
$^*$ University of Minnesota,
Minneapolis, MN 55404, USA, 
Email: cardo089@umn.edu   
\vspace{-1em}
 \thanks{The work of A. Dytso and H. V. Poor was supported in part by the U. S. National Science Foundation under Grant CCF-1513915. The work of M. Cardone was supported in part by the U.S. National Science Foundation under Grant CCF-1849757. 

The results of this paper were accepted in part for publication at the 2019 IEEE International Symposium on Information Theory.}

}

\maketitle
\begin{abstract}
This paper proposes an estimation framework to assess the performance of sorting over perturbed/noisy data.
In particular, the recovering accuracy is measured in terms of Minimum Mean Square Error (MMSE) between the values of the sorting function computed on data without perturbation and the estimator that operates on the sorted noisy data.
It is first shown that, under certain symmetry conditions, satisfied for example by the practically relevant Gaussian noise perturbation, the optimal estimator can be expressed as a linear combination of estimators on the unsorted data.
Then, two suboptimal estimators are proposed and performance guarantees on them are derived with respect to the optimal estimator.
Finally, some surprising properties on the MMSE of interest are discovered.
For instance, it is shown that the MMSE grows sublinearly with the data size, and that commonly used MMSE lower bounds such as the Bayesian Cram\'er-Rao and the maximum entropy bounds either cannot be applied or are not suitable.
\end{abstract}

\section{Introduction}
Today, the sorting function is widely used in a  broad variety of applications. For instance, it is well recognized that sorting is a benchmark for several modern recommender and distributed computing systems (e.g., Hadoop MapReduce).

In this paper, we focus on analyzing the performance of the sorting function over data that is perturbed. 
Such a data perturbation might occur because of data privacy purposes, i.e., the data is sensitive (e.g., clinical/genomic health) and hence it is required to remain confidential/private. 
This gives rise to a natural question: {\em How does data perturbation affect the performance of the sorting function?}

Our main objective is to propose and make first steps in analyzing an estimation framework, which seeks to answer the question above. Towards this end, we discover several properties that shed light on the effect of the sorting function on the data statistics.

\subsection{Related Work}
In this work, we leverage tools and properties from the theory of order statistics. 
Order statistics indeed represents an indispensable tool in the modern theory of statistical inference. Next, we briefly mention some notable examples.  

The estimation of  parameters of a family of  distributions from an ordered vector has received a considerable attention.   
For example, for a location-scale family of  distributions with density function $\frac{1}{\sigma}  f \left(  \frac{x-\mu}{\sigma }\right)$, the best linear unbiased estimator (BLUE) of $\sigma$ and $\mu$ has been found in~\cite{lloyd1952least}.  In particular, the BLUE was shown to be a function of only the covariance  matrix and  the mean vector of the order statistics.  However, note that the computation of the covariance matrix and the mean vector of  the order statistics is often a formidable  task.  Explicit expressions for moments of  order statistics  are in fact known only for some specific distributions and have been tabulated in~\cite{harter1996crc}.  A rich body of literature also exists on universal bounds on moments of order statistics~\cite{david2004order}. 

Having observed a  partial sample of the first $r$ order statistics from a sample of size $n$, the best linear unbiased predictor (BLUP) of the remaining  $n-r$ terms has been characterized in~\cite{goldberger1962best}.  Specifically, the predictor in~\cite{goldberger1962best} was shown to depend only on the covariance matrix and the mean  of the order statistics. Interesting connections between the BLUE and BLUP have been derived in~\cite{doganaksoy1997useful}. 

Due to its application to life-testing experiments,  inference of censored (i.e., incomplete data) order statistics  has also received attention.  Interestingly, for various  types of censoring protocols,  closed-form expressions for maximum likelihood estimators (MLEs) of parameters  are available.  A comprehensive survey of several censoring scenarios can be found 
in~\cite{bain2017statistical}.  

Order statistics also  appears in the study of outliers since  these are expected to be a few extreme order statistics.  Several effective tests are formed from extreme order statistics that seek to compute the deviation of the candidate outliers from the rest of the data  \cite{barnett1974outliers}.  Another application of order statistics is on  the goodness-of-fit tests. The most classical example of such a test is the  Shapiro and Wilk's test for normality~\cite{shapiro1965analysis}. 

 A collection of articles summarizing applications, results  and  historic perspectives on order statistics can be found in~\cite{balakrishnan1998handbook}. 
    
While classical works have focused on estimating parameters of distributions from order statistics or recovering incomplete ordered sequences, modern applications  have focused on  sorting with noisy data. 
 In~\cite{pananjady2017denoising}, the authors considered an additive noise linear regression model in which the output variables are permuted and might be incorrectly associated with the input variables. The goal is to estimate the correct permutation matrix and the linear parameter.  Further extensions of~\cite{pananjady2017denoising} include~\cite{rigollet2018uncoupled} where the authors considered an additive noise regression model with  the goal of estimating an unknown  monotone increasing function of the input variables. 

There is also a large body of works in computer science that focuses on sorting with unreliable machines that perform pairwise comparisons; see~\cite{braverman2008noisy, mao2017minimax} and references therein.

Different from the aforementioned works, we are here interested in estimating the values of the components of the original sorted vector, by performing joint estimation and sorting. Moreover, we  focus on the Bayesian setting, i.e., we assume a prior distribution on the input variables. This work builds on our preliminary results in~\cite{ISIT2019}, and extends them 
by studying the effect of sorting on the data statistics.

    \begin{figure*}[h!]
        \centering
        \includegraphics[width=0.79\textwidth]{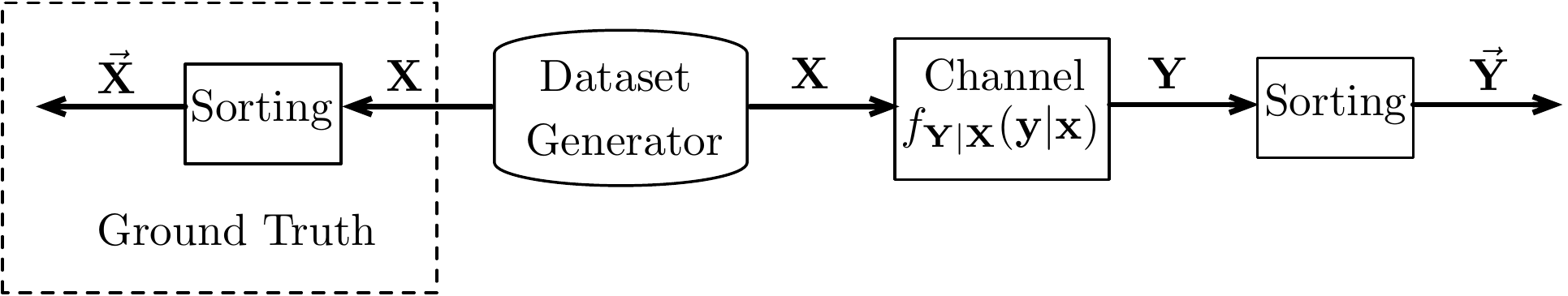}
        \caption{Graphical representation of the proposed framework.}
        \label{fig:Framework}
    \end{figure*}

\subsection{Contributions}

In this work, we focus on assessing the performance of the sorting function over data that is perturbed.
Our goal is to quantify the performance loss of the sorting function versus different levels of noise perturbation.
Our main contributions can be summarized as follows:
\begin{enumerate}[leftmargin=*]
\item We pose the problem within an {\em estimation framework}. In particular, as a metric, we adopt the Minimum Mean Square Error (MMSE) between the values of the sorting function computed on the original data (i.e., with no perturbation) and the estimation that uses the sorting function applied to the noisy version of the data. Our proposed framework is presented in Section~\ref{sec:SysModel}.
\item We analyze the {\em optimal estimator}, i.e., the conditional expectation of observing the original data as sorted, given the observation of the noisy sorted data. 
We show that, under certain  symmetry conditions, satisfied for example by the practically relevant Gaussian noise perturbation, the optimal estimator can be expressed as a linear combination of estimators of the unsorted data from unsorted noisy  observations. 
Our analysis on the optimal estimator can be found in Section~\ref{sec:OptEstimator}. 
\item We leverage the structure of the optimal estimator to propose two {\em suboptimal estimators}, and we analyze their performance with respect to the optimal estimator. 
We prove that, with Gaussian statistics, one estimator is asymptotically optimal in the low noise regime, whereas the other is asymptotically optimal in the high noise regime.
We also characterize the structure of the MLE, and we numerically assess its performance with Gaussian statistics.
Our analysis on the suboptimal estimators can be found in Section~\ref{sec:SubEst} and Section~\ref{sec:GaussianStat}.
\item We discover several {\em properties on the MMSE} of interest, which highlight interesting observations on how the sorting function affects the statistics.
For instance, in Section~\ref{sec:HighN} we show that for order statistics the MMSE grows sublinearly with the size of the vectors; this represents a fundamental difference from unsorted data for which the MMSE grows linearly in the size of the vector. 
Another surprising observation concerns lower bounds on the MMSE of interest. In Section~\ref{sec:LowerBoundsOnMMSE}, we indeed show that the Bayesian Cram\'er-Rao bound cannot be applied, and the maximum entropy bound does not scale properly with the size of the~vectors.  
\end{enumerate}



\section{Notation, Definitions and System Model}
\label{sec:SysModel}

Boldface upper case letters $\mathbf{X}$ denote vector random variables; the boldface lower case letter $\bf{x}$ indicates a specific realization of $\mathbf{X}$; $\vec{\mathbf{X}}$ is $\mathbf{X}$ ordered in ascending order; $X_{i}$ specifies the $i$-th entry of ${\mathbf{X}}$ and $X_{(i)}$ indicates the $i$-th entry of $\vec{\mathbf{X}}$;  
$\|\mathbf{x}\|$ denotes the $\ell_2$-norm of the vector $\mathbf{x}$;   $\mathbf{I}_n$ is the identity matrix of dimension $n$; $\mathbf{0}_n$ is the column vector of dimension $n$ of all zeros; $\delta(x)$ is the Dirac delta function; $1_{ \{ S\}}(x_i)$ is the indicator function over the set $S$; $[n]$ indicates the set of integers $\{1,\dots,n\}$. 
The set $\vec{\mathbb{R}}_n= \{ {\bf x} :  \vec{\bf x} \in \mathbb{R}_n \}$;   $\mathcal{P}$ denotes the collection of all permutations of the elements of $[n]$ and  $| \mathcal{P}| =n!$. 
$\mathbf{P}_ \pi $ where $\pi \in \mathcal{P}$  is a permutation matrix associated with $\pi$.  For example,  for $n=3$
\begin{equation*}
\mathbf{P}_ { \{1,2,3 \}}= \mathbf{I}_3,
\qquad
\text{and} 
\qquad
\mathbf{P}_ { \{2,3,1 \} }=  \left[ \begin{array}{ccc}  0 & 1  & 0 \\ 0 & 0 & 1\\ 1 & 0 & 0\end{array}  \right]. 
\end{equation*}

\smallskip

We next provide a definition, which will be used in the proof of our main results.
\begin{defin}
\label{def:ExchRanVar}
A sequence of random variables $U_1, U_2, \ldots, U_n$ is said to be exchangeable if 
for any permutation $(\pi_1,\pi_2,\ldots,\pi_n)$ of the indices $[n]$, we have that
\begin{equation*}
(U_1,U_2,\ldots,U_n) \stackrel{d}{=} (U_{\pi_1}, U_{\pi_2}, \ldots, U_{\pi_n}),
\end{equation*}
where $\stackrel{d}{=}$ denotes equality in distribution.
\end{defin}

 Note that any convex combination or mixture distribution of independent and identically distributed sequences of random variables is exchangeable.

\smallskip

We consider the framework shown in Fig.~\ref{fig:Framework}, where an $n$-dimensional random vector $\mathbf{X}$
is generated according to a certain probability density function (PDF) $f_{\mathbf{X}}(\cdot)$ and then passed through a noisy channel with a transition PDF equal to $f_{\mathbf{Y}|\mathbf{X}}(\cdot | \cdot)$.
This is assumed to be a {\it parallel} channel, i.e.,
\begin{equation*}
f_{\mathbf{Y}|\mathbf{X}}(\mathbf{y}|\mathbf{x}) = \prod_{i=1}^n f_{{Y}_i|{X}_i} (y_i|x_i).
\end{equation*}
The output of the channel -- denoted as $\mathbf{Y}$ -- is finally sorted in ascending order, i.e.,$\vec{\mathbf{Y}}$ is the sorted version of $\mathbf{Y}$.

In this work, we are interested in characterizing the MMSE 
of 
estimating $\vec{\mathbf{X}}$ -- which denotes the sorted version of ${\mathbf{X}}$ 
(i.e., the ground truth) -- when $\vec{\mathbf{Y}}$ is observed.

\section{On the Optimal MMSE Estimator}
\label{sec:OptEstimator}
The objective of this section  is to  study the MMSE of estimating $\vec{\mathbf{X}}$ from an observation $\vec{\mathbf{Y}}$. 
The MMSE is given~by
\begin{equation}
\mmse(\vec{\mathbf{X}}|\vec{\mathbf{Y}}) = \mathbb{E} \left [ \left \| \vec{\mathbf{X}} - \mathbb{E} \left [ \vec{\mathbf{X}} | \vec{\mathbf{Y}}\right ] \right \|^2 \right ].
\label{eq:MMSE}
\end{equation}
It is well  known that the conditional expectation $\mathbb{E} \left [ \vec{\mathbf{X}} | \vec{\mathbf{Y}}\right ] $  is the optimal estimator under the square error  criterion. The next theorem provides a characterization of $\mathbb{E} \left [ \vec{\mathbf{X}} | \vec{\mathbf{Y}}\right ] $  in terms of the distribution of  $(\X,\Y)$ under certain symmetry conditions.   

\begin{thm}
\label{thm:OptimalEst}
Let $(\X,\Y)$ be continuous random vectors.   Assume that  $\X$ is exchangeable 
and that, $\forall (\mathbf{x},\mathbf{y})$
\begin{align}
\mathbb{E}_{\Pi_1,\Pi_2} \left [ f_{\mathbf{Y}|\mathbf{X}}(\mathbf{P}_{\Pi_2} {\mathbf{y}}|\mathbf{P}_{\Pi_1}{\mathbf{x}}) \right ] 
= \mathbb{E}_{\Pi} \left [ f_{\mathbf{Y}|\mathbf{X}}(\mathbf{P}_{\Pi} \mathbf{y}|\mathbf{x}) \right ],   \label{eq:PermuationSymmetryCondition}
\end{align}
where $(\Pi, \Pi_1, \Pi_2)$ are  mutually independent and uniformly distributed on $\mathcal{P}$. 
Then, 
\begin{align}
& \frac{\mathbb{E}[ \vec{\mathbf{X}} | \vec{\mathbf{Y}} \!=\! \vec{\mathbf{y}}] }{n!}
\! =\!  \!\sum_{ \pi \in  \mathcal{P} } \!\!\frac{ f_{{\mathbf{Y}} }( \mathbf{P}_{\pi} \vec{\mathbf{y}})}{f_{\vec{\mathbf{Y}} }(\vec{\mathbf{y}})}   \mathbb{E} \left[ \X  \cdot 1_{\vec{\mathbb{R}}_n}(\mathbf{X}) |  \mathbf{Y}\!=\! \mathbf{P}_{\pi}\vec{\mathbf{y}}    \right] .    \label{eq:EstimatorOnlyXsymmetric} 
\end{align}
In addition, if $\Y$ is exchangeable,
then \begin{align}
 \mathbb{E}[\vec{\mathbf{X}}| \vec{\mathbf{Y}}=\vec{\mathbf{y}}] 
 &=   \sum_{ \pi \in  \mathcal{P} }  \mathbb{E} \left[ \X  \cdot 1_{\vec{\mathbb{R}}_n}(\mathbf{X}) |  \mathbf{Y}= \mathbf{P}_{\pi}\vec{\mathbf{y}}    \right] . 
 \label{eq:EstimatorBothXandYsymmetric} 
\end{align} 
\end{thm}
\begin{IEEEproof}
 We have
\begin{align*}
&\mathbb{E} \left  [X_{(k)}| \vec{\mathbf{Y}} = \vec{\mathbf{y}}  \right]  \notag\\
=&    \int_{ \vec{\mathbb{R}}_n}  t_k   f_{\vec{\mathbf{X}}|\vec{\mathbf{Y}} }( \mathbf{t} | \vec{\mathbf{y}})   {\rm d} \mathbf{t} \\
   \stackrel{{\rm{(a)}}}{=}  &   \int_{ \vec{\mathbb{R}}_n}  t_k    \frac{n!}{f_{\vec{\mathbf{Y}} }(\vec{\mathbf{y}})}  f_{\mathbf{X}}(\mathbf{t})
\sum_{ \pi \in  \mathcal{P} }  f_{\mathbf{Y}|\mathbf{X} } (\mathbf{P}_{\pi}\vec{\mathbf{y}} | \mathbf{t})  {\rm d} \mathbf{t} \\
%
= &     \frac{n!}{f_{\vec{\mathbf{Y}} }(\vec{\mathbf{y}})} 
\sum_{ \pi \in  \mathcal{P} }  \int_{ \mathbb{R}_n}  t_k \cdot 1_{ \vec{\mathbb{R}}_n }(\mathbf{t})   f_{\mathbf{Y}|\mathbf{X} } (\mathbf{P}_{\pi}\vec{\mathbf{y}} | \mathbf{t})  f_{\mathbf{X}}(\mathbf{t})  {\rm d} \mathbf{t} \\
= &  \frac{n!}{f_{\vec{\mathbf{Y}} }(\vec{\mathbf{y}})}  \sum_{ \pi \in  \mathcal{P} }  \frac{f_{{\mathbf{Y}} }(\mathbf{P}_{\pi}\vec{\mathbf{y}})}{f_{{\mathbf{Y}} }(\mathbf{P}_{\pi}\vec{\mathbf{y}})}  \int_{\mathbb{R}_n}      t_k  1_{\vec{\mathbb{R}}_n}(\mathbf{t})   f_{\mathbf{X}}(\mathbf{t}) \notag\\
& \quad \cdot 
  f_{\mathbf{Y}|\mathbf{X} } (\mathbf{P}_{\pi}\vec{\mathbf{y}} | \mathbf{t})    {\rm d} \mathbf{t}\\
 \stackrel{{\rm{(b)}}}{=} &    \frac{n!}{f_{\vec{\mathbf{Y}} }(\vec{\mathbf{y}})} \sum_{ \pi \in  \mathcal{P} }  f_{{\mathbf{Y}} }(\mathbf{P}_{\pi}\vec{\mathbf{y}})  \int_{\mathbb{R}_n}      t_k  1_{\vec{\mathbb{R}}_n}(\mathbf{t})   f_{\mathbf{X}| \mathbf{Y}}( \mathbf{t}|\mathbf{P}_{\pi}\vec{\mathbf{y}} )    {\rm d} \mathbf{t}\\
    = & \frac{n!}{f_{\vec{\mathbf{Y}} }(\vec{\mathbf{y}})} \sum_{ \pi \in  \mathcal{P} }  f_{{\mathbf{Y}} }(\mathbf{P}_{\pi}\vec{\mathbf{y}})   \mathbb{E} \left[ X_k  \cdot 1_{\vec{\mathbb{R}}_n}(\mathbf{X}) |  \mathbf{Y}=\mathbf{P}_{\pi}\vec{\mathbf{y}}    \right],
  \end{align*}
  where the labeled equalities follow from:  $\rm{(a)}$ the identity
\begin{equation*}
f_{\vec{\mathbf{X}}|\vec{\mathbf{Y}} }( \vec{\mathbf{x}} | \vec{\mathbf{y}}) = \frac{n!}{f_{\vec{\mathbf{Y}} }(\vec{\mathbf{y}})}  f_{\mathbf{X}}(\vec{\mathbf{x}})
\sum_{ \pi \in  \mathcal{P} }  f_{\mathbf{Y}|\mathbf{X} } (\mathbf{P}_{\pi}\vec{\mathbf{y}} | \vec{\mathbf{x}})
\end{equation*}
 in Lemma~\ref{lem:forMAP} in Appendix~\ref{app:AuxRes};  and $\rm{(b)}$ using Bayes' rule.
This concludes the proof of \eqref{eq:EstimatorOnlyXsymmetric}. 
To show that~\eqref{eq:EstimatorBothXandYsymmetric} holds observe that, if ${\bf Y}$ is exchangeable,
then  
\begin{equation}
\frac{ f_{{\mathbf{Y}} }(\mathbf{P}_{\pi}\vec{\mathbf{y}})}{f_{\vec{\mathbf{Y}} }(\vec{\mathbf{y}})}  = \frac{ f_{{\mathbf{Y}} }(\vec{\mathbf{y}})}{f_{\vec{\mathbf{Y}} }(\vec{\mathbf{y}})} = \frac{1}{n!},
\end{equation}
where the last equality follows by using Lemma~\ref{lem:Lemma_exchangeable} in Appendix~\ref{app:AuxRes}.
%
%
This concludes the proof of Theorem~\ref{thm:OptimalEst}.
\end{IEEEproof}

%
%
%
%

\begin{rem}
 Examples of noisy transformations that satisfy~\eqref{eq:PermuationSymmetryCondition} include the practically relevant Gaussian distribution.
\end{rem}

 Theorem~\ref{thm:OptimalEst} allows to express the optimal estimator  of elements of $\vec{\mathbf{X}}$ from  an observation $\vec{\mathbf{Y}}$   as a linear combination of estimators of the original  unsorted  $\X$ from the unsorted channel observation $\Y$.  This has the potential  of significantly simplifying the  computation of the conditional expectation as there is no need to find the joint distribution of  $(\vec{\mathbf{X}}, \vec{\mathbf{Y}})$. 

In the remaining part of the paper, we assume that $\X$ and $\Y$ are both exchangeable, and hence we focus on further analyzing the optimal estimator in~\eqref{eq:EstimatorBothXandYsymmetric}.

\section{Suboptimal Estimators}
\label{sec:SubEst}
Observe that the conditional expectation in~\eqref{eq:EstimatorBothXandYsymmetric} can be equivalently written as follows:
\begin{align}
 &\mathbb{E} \left[ X_k  \cdot 1_{\vec{\mathbb{R}}_n}(\mathbf{X}) |  \mathbf{Y}=\mathbf{P}_{\pi}\vec{\mathbf{y}}    \right]  \notag\\
 =&   \mathbb{E} \left[ X_k     | \mathbf{X} \in \vec{\mathbb{R}}_n,      \mathbf{Y}=\mathbf{P}_{\pi}\vec{\mathbf{y}}    \right]     \mathbb{P} \left[  \mathbf{X} \in \vec{\mathbb{R}}_n |  \mathbf{Y}=\mathbf{P}_{\pi}\vec{\mathbf{y}}      \right] .  \label{eq:IndicatorFunctionConditioning}
\end{align}
We  now use the identity above to propose and study two different suboptimal estimators. The idea is to approximate~\eqref{eq:IndicatorFunctionConditioning} by dropping certain dependencies  on $ \mathbf{Y}$ or $\mathbf{X} \in \vec{\mathbb{R}}_n$. Moreover, we also discuss the structure of the MLE. 

The first suboptimal estimator of $X_{(k)}$ that we analyze is
\begin{subequations}
\label{eq:lowNoiseEstiator}
\begin{equation}
\hat{f}_k( \vec{\mathbf{y}}) \!=\!  \sum_{ \pi \in  \mathcal{P} }    \mathbb{E} \left[ X_k   |       \mathbf{Y}\!=\!\mathbf{P}_{\pi}\vec{\mathbf{y}}    \right] \cdot    \mathbb{P} \left[  \mathbf{X} \in \vec{\mathbb{R}}_n |  \mathbf{Y}\!=\!\mathbf{P}_{\pi}\vec{\mathbf{y}}      \right] , \label{eq:ProposeSubOptimalEstimator}
\end{equation}
for $k \in [n],$
and  the following suboptimal estimator of $\vec{\mathbf{X}}$
\begin{equation}
 {\bf  \hat{f}}( \vec{\mathbf{y}}) = [\hat{f}_1( \vec{\mathbf{y}}), \ldots, \hat{f}_n( \vec{\mathbf{y}})].
\end{equation}
\end{subequations}
Observe that the only difference between the optimal estimator in~\eqref{eq:IndicatorFunctionConditioning} and the  suboptimal estimator in \eqref{eq:ProposeSubOptimalEstimator} is that in the latter the conditioning on $ \mathbf{X} \in \vec{\mathbb{R}}_n$ has been dropped. In other words, we are implicitly using the approximation   $ \mathbb{E} \left[ X_k     | \mathbf{X} \in \vec{\mathbb{R}}_n,      \mathbf{Y}  \right]  \approx  \mathbb{E} \left[ X_k     |      \mathbf{Y}    \right]  $.

The next theorem,  whose proof can be found in Appendix~\ref{app:proof:thm:ErrorForApproximateEstiamtor}, compares the performance of the optimal estimator in \eqref{eq:EstimatorBothXandYsymmetric} to the proposed suboptimal estimator in \eqref{eq:ProposeSubOptimalEstimator}. 
\begin{thm}\label{thm:ErrorForApproximateEstiamtor}  Assume that $\X$ and $\Y$ are exchangeable and that  the assumption in \eqref{eq:PermuationSymmetryCondition} holds. Let
\begin{equation}
  \Delta=\left| \mathbb{E} \left [ \left \| \vec{\mathbf{X}} - {\bf  \hat{f}}( \vec{\mathbf{Y}}) \right \|^2 \right ]-  \mmse(\vec{\mathbf{X}}|  \vec{\mathbf{Y}}) \right|  .  \label{eq:approximationError}
\end{equation}
Then,
\begin{equation}
 \Delta \! \le \! \Delta_{\text{up}} = \sum_{ \pi \in  \mathcal{P} }     \mathbb{E}  \left[  \|\X\|^2   g \left(\mathbf{P}_{\pi}\mathbf{Y} \right) \mid     \mathbf{Y} \in \vec{\mathbb{R}}_n \right],\label{eq:GapToOptimalityOfSubOptmalEstimator}
\end{equation}
where
\begin{equation}
\hspace{-0.1cm} g({\bf y}) = \mathbb{P}  \hspace{-0.05cm}\left[  \mathbf{X} \in \vec{\mathbb{R}}_n |  \mathbf{Y}=\mathbf{y}     \right]   \hspace{-0.1cm}\left(1-\mathbb{P} \left[  \mathbf{X} \in \vec{\mathbb{R}}_n |  \mathbf{Y}=\mathbf{y}     \right]  \right)	. 
\end{equation}
\end{thm} 

Intuitively, the estimator in \eqref{eq:ProposeSubOptimalEstimator} performs well when $\X$ and $ \Y $ have a high degree of correlation, and the information contained into $\Y$ is enough to produce a good estimate of $\vec{\X}$.  This scenario typically occurs when the noise between $\X$ and $\Y$ is weak. However, this estimator starts performing poorly when $\X$ and $ \Y $ are only weakly dependent on each other.  In this case, in fact, ignoring the condition $\mathbf{X} \in \vec{\mathbb{R}}_n$ can lead to a severe penalty. Motivated by this observation, we next propose an estimator that MMSE wise works well in the opposite regime, that is when $\X$ and $\Y$ are weakly dependent.  

The second suboptimal estimator of $X_{(k)}$ that we analyze~is
\begin{subequations}
\label{eq:highNoiseRegime}
\begin{equation}
\hat{h}_k( \vec{\mathbf{y}}) \!=\! \sum_{ \pi \in  \mathcal{P} } \E[ X_k | \X \!\in \! \vec{\mathbb{R}}_n  ] \cdot \mathbb{P}[ \X \in  \vec{\mathbb{R}}_n | \mathbf{Y}\!=\!\mathbf{P}_{\pi}\vec{\bf y}    ],
\label{eq:highNoiseRegimea}
\end{equation}
for $k \in [n]$, 
and  the following suboptimal estimator of $\vec{\mathbf{X}}$
\begin{equation}
\label{eq:SubEstHighNoise}
 {\bf  \hat{h}}( \vec{\mathbf{y}}) = [\hat{h}_1( \vec{\mathbf{y}}), \ldots, \hat{h}_n( \vec{\mathbf{y}})].
\end{equation}
\end{subequations}
Observe that the only difference between the optimal estimator in~\eqref{eq:IndicatorFunctionConditioning} and the  suboptimal estimator in~\eqref{eq:highNoiseRegimea} is that in the latter the conditioning on $ \mathbf{Y}$ inside the conditional expectation has been dropped. In other words, we are implicitly using $ \mathbb{E} \left[ X_k     | \mathbf{X} \in \vec{\mathbb{R}}_n,      \mathbf{Y}  \right]  \approx  \mathbb{E} \left[ X_k     |     \mathbf{X} \in \vec{\mathbb{R}}_n    \right]  $.
Before analyzing the performance of the suboptimal estimator given in~\eqref{eq:highNoiseRegime}, we state the next lemma (whose proof is provided in Appendix~\ref{app:EstHighNoise}), which offers a simplified expression for the estimator in~\eqref{eq:highNoiseRegime}. 
\begin{lem}
\label{lem:EquivalentHighNoise}
Assume that $\X$ and $\Y$ are exchangeable random vectors.
Then, the estimator in~\eqref{eq:highNoiseRegime} can be written as
\begin{align}
\label{eq:EquivalentHighNoise}
\hat{h}_k( \vec{\mathbf{y}}) &= \E[ X_k | \X \in  \vec{\mathbb{R}}_n  ]  = \E[X_{(k)}] , \ \forall k \in [n].
\end{align}
\end{lem}
In other words, Lemma~\ref{lem:EquivalentHighNoise} shows that by dropping the conditioning on  ${\mathbf{Y}}$ inside the conditional expectation in the optimal estimator in~\eqref{eq:IndicatorFunctionConditioning}, then the resulting suboptimal estimator only uses the knowledge of the prior distribution of $\vec{\mathbf{X}}$.   The optimal MSE of such an estimator can be easily computed as
\begin{align}
\label{eq:MSESubHighNoise}
\mathbb{E} \left [ \left \| \vec{\mathbf{X}} - {\bf  \hat{h}}( \vec{\mathbf{Y}}) \right \|^2 \right ] = \mathbb{E} \left [ \left \| \vec{\mathbf{X}} - \mathbb{E}[\vec{\X}] \right \|^2 \right ] = \mathsf{Var}(\vec{\mathbf{X}}). 
\end{align}
Intuitively, such an estimator performs well when the output $\Y$ contains little information about the input $\X$, e.g., when the noise between $\X$ and $\Y$ is high. Sufficient conditions for 
the MSE optimality of this estimator are given in the next  
theorem, whose proof can be found in Appendix~\ref{app:PerfHighNoiseEst}.

\begin{thm}\label{thm:HighSigmaRegimeGeneral}  Assume that  for every $t \in \mathbb{N}$, $\X$  and $\Y_t$  are exchangeable random vectors, and that the assumption in~\eqref{eq:PermuationSymmetryCondition} holds. In addition, assume that
\begin{align}
 \lim _{t \to \infty} \E[\X | \Y_t] = \E[\X ] \text{ a.s.}  \label{eq:ConitinoutyAssumption}
 \end{align}
 and let 
\begin{align}
\Delta_t=\left| \mathbb{E} \left [ \left \| \vec{\mathbf{X}} - {\bf  \hat{h}}( \vec{\mathbf{Y}}_t) \right \|^2 \right ]-  \mmse(\vec{\mathbf{X}}|  \vec{\mathbf{Y}}_t) \right|  .   \label{eq:ErrorHighSnr}
\end{align} 
Then, if $\E[\| \X\|^2] < \infty$, we have that
\begin{align}
\lim_{t \to \infty} \Delta_t=0. 
\end{align}
\end{thm} 
\begin{rem} The assumption in~\eqref{eq:ConitinoutyAssumption} holds for several channels such as the practically relevant Gaussian noise channel.
This case will be studied in more detail in Section~\ref{sec:GaussianStat}. 
Another important feature of Theorem~\ref{thm:HighSigmaRegimeGeneral} is that the conditions for the convergence depend only on $(\X,\Y)$  and not on $(\vec{\mathbf{X}},\vec{\mathbf{Y}}).$
\end{rem}
We now conclude this section by characterizing the structure of the MLE of $\vec{\mathbf{X}}$ from $\vec{\mathbf{Y}}$ in terms of the conditional distribution of the original unsorted pair $(\X,\Y)$. By leveraging Lemma~\ref{lem:forMAP} in Appendix~\ref{app:AuxRes}, we obtain
\begin{equation}
 {\bf  \hat{\boldsymbol{\ell}}}_{\text{MLE}}( \vec{\mathbf{y}})  \in  \arg\max_{ {\bf t} \in   \vec{\mathbb{R}}_n }    \sum_{ \pi \in  \mathcal{P} }  f_{\mathbf{Y}|\mathbf{X} } (\mathbf{P}_{\pi}\vec{\mathbf{y}} | \mathbf{t}).  \label{eq:MaximumLikelihood_Estimator}
 \end{equation}

\section{Suboptimal Estimators Evaluated with Gaussian Statistics}
\label{sec:GaussianStat}

In this section, we consider the practically relevant case of Gaussian noise, i.e., we assume that  $\Y| \X=\mathbf{x} ~\sim \mathcal{N}({\bf x}, \sigma^2 \mathbf{I}_n)$. In particular, we assess the performance of the two suboptimal estimators in~\eqref{eq:lowNoiseEstiator} and in~\eqref{eq:highNoiseRegime}, as well as of the MLE in~\eqref{eq:MaximumLikelihood_Estimator}, by comparing them with the optimal estimator in Theorem~\ref{thm:OptimalEst}.

\begin{figure*}[h!]
\centering
\begin{subfigure}[t]{0.45\textwidth}
%
%
\pgfplotsset{every axis plot/.append style={very thick}}

%

\begin{tikzpicture}

\begin{axis}[%
width=9cm,
height=6.5cm,
at={(1.73in,1.152in)},
xmin=0,
xmax=10,
xlabel style={font=\color{white!15!black}, below=-2mm},
xlabel={$\sigma$},
ymin=0,
ymax=2.5,
ylabel style={font=\color{white!15!black}},
ylabel style={at={(axis description cs:0.09,.5)},anchor=south},
ylabel={MSE},
axis background/.style={fill=white},
xmajorgrids,
ymajorgrids,
legend style={at={(0.98,0.42)}, align=left, draw=white!15!black, font=\footnotesize}
]
\addplot [color=black, line width=2.0pt]
  table[row sep=crcr]{%
0	3.77475828372553e-15\\
0.25	0.108566683336917\\
0.5	0.340422133498652\\
0.75	0.568205335661287\\
1	0.746209583295326\\
1.25	0.87727310588897\\
1.5	0.974742900578599\\
1.75	1.04855660852357\\
2	1.10459706844141\\
2.25	1.1479891796672\\
2.5	1.18210355021605\\
2.75	1.20897908384781\\
3	1.23058336495719\\
3.25	1.247899711517\\
3.5	1.2627103783219\\
3.75	1.27446249795177\\
4	1.28415859348568\\
4.25	1.29240008875877\\
4.5	1.29989617668168\\
4.75	1.30625456299558\\
5	1.3114998456096\\
5.25	1.31610054447978\\
5.5	1.31971531923066\\
5.75	1.32324690243772\\
6	1.32656774412599\\
6.25	1.32934586339629\\
6.5	1.33206544859391\\
6.75	1.33404109576892\\
7	1.33618754818678\\
7.25	1.33784521466268\\
7.5	1.33959524987517\\
7.75	1.34120429630327\\
8	1.34215320553204\\
8.25	1.34366939278216\\
8.5	1.34460990737459\\
8.75	1.34553444281948\\
9	1.34660497471842\\
9.25	1.34729180600529\\
9.5	1.34816542582006\\
9.75	1.34896912887791\\
10	1.34980539412891\\
20	1.35956698137311\\
30	1.36170387372467\\
40	1.36238228485336\\
50	1.36271318802632\\
};
\addlegendentry{$\mmse(\vec{\mathbf{X}}|\vec{\mathbf{Y}})$}

\addplot [color=black, dashdotted, line width=2.0pt]
  table[row sep=crcr]{%
0	-1.86517468137026e-14\\
0.25	0.128121300118725\\
0.5	0.459045479972726\\
0.75	0.829687344385138\\
1	1.12009423760621\\
1.25	1.3195588395884\\
1.5	1.45820698561308\\
1.75	1.55920536406095\\
2	1.63577022950629\\
2.25	1.6953344756684\\
2.5	1.74202082093123\\
2.75	1.77924341440589\\
3	1.80959683980155\\
3.25	1.83418075861414\\
3.5	1.85432726878616\\
3.75	1.87154739237236\\
4	1.88551807067447\\
4.25	1.89756649143298\\
4.5	1.90795906085041\\
4.75	1.91707184403165\\
5	1.92421087237117\\
5.25	1.9309973891014\\
5.5	1.93705455277997\\
5.75	1.94191948293493\\
6	1.94679144263778\\
6.25	1.95039540435787\\
6.5	1.95385517282409\\
6.75	1.95743614988634\\
7	1.96041328231923\\
7.25	1.96285238455922\\
7.5	1.96546114189751\\
7.75	1.96751548593941\\
8	1.96933948950096\\
8.25	1.9710352010124\\
8.5	1.97296997951738\\
8.75	1.97418133728611\\
9	1.97560511159906\\
9.25	1.97694236448774\\
9.5	1.9781171275047\\
9.75	1.97918216359139\\
10	1.98010689663402\\
20	1.99495659942453\\
30	1.99764405814211\\
40	1.99859173755407\\
50	1.99905657082254\\
};
\addlegendentry{ \text{MSE} of ${\bf  \hat{f}}$}

\addplot [color=black, dotted,  line width=2.0pt]
  table[row sep=crcr]{%
0	1.363\\
0.25	1.363\\
0.5	1.363\\
0.75	1.363\\
1	1.363\\
1.25	1.363\\
1.5	1.363\\
5	1.363\\
5.25	1.363\\
7	1.363\\
7.25	1.363\\
8.25	1.363\\
9.25	1.363\\
9.75	1.363\\
10	1.363\\
20	1.363\\
30	1.363\\
40	1.363\\
50	1.363\\
};
\addlegendentry{$\mathsf{Var}(\vec{\mathbf{X}})$ or MSE of $ {\bf  \hat{h}}$}

\addplot [color=black, dashed, line width=2.0pt]
  table[row sep=crcr]{%
0.2	0.0826327233965845\\
0.25	0.128001322605882\\
0.3	0.181722095680013\\
0.35	0.246788099801246\\
0.4	0.322466143512432\\
0.45	0.411025634407546\\
0.5	0.498939819007856\\
0.55	0.604076791053801\\
0.6	0.724036739141245\\
0.65	0.836969779355409\\
0.7	0.957358539446133\\
0.75	1.08723921847115\\
0.8	1.23393943211941\\
0.85	1.37102898167028\\
0.9	1.50915582484377\\
0.95	1.6680845067921\\
1	1.88200501998448\\
1.05	2.05460368471022\\
1.1	2.2488507179226\\
1.15	2.40341388193628\\
1.2	2.62439596136144\\
1.25	2.81906189196565\\
1.3	3.06276802622993\\
1.35	3.24278047117031\\
1.4	3.48763922220198\\
1.45	3.68456006537547\\
1.5	3.92143656147965\\
}; 
\addlegendentry{MSE of MLE}

\end{axis}  
\end{tikzpicture}%
\caption{$n=2$}
\end{subfigure}\quad \quad
\begin{subfigure}[t]{0.45\textwidth}
%
%
\pgfplotsset{every axis plot/.append style={very thick}}

\definecolor{mycolor1}{rgb}{0.00000,0.45000,0.74000}%

\begin{tikzpicture}

\begin{axis}[%
width=9cm,
height=6.5cm,
at={(1.73in,1.152in)},
xmin=0,
xmax=10,
xlabel style={font=\color{white!15!black},below=-2mm},
xlabel={$\sigma$},
ymin=0,
ymax=3.5,
ylabel style={font=\color{white!15!black}},
ylabel style={at={(axis description cs:0.09,.5)},anchor=south},
ylabel={MSE},
axis background/.style={fill=white},
xmajorgrids,
ymajorgrids,
legend style={at={(0.98,0.41)}, align=left, draw=white!15!black, font=\footnotesize}
]

\addplot [color=black, line width=2.0pt]
  table[row sep=crcr]{%
0	-3.5527136788005e-15\\
0.25	0.151715339286685\\
0.5	0.44379298970622\\
0.75	0.702748230917135\\
1	0.892826655023399\\
1.25	1.03089369005599\\
1.5	1.13464806076762\\
1.75	1.21437216297872\\
2	1.27625826724762\\
2.25	1.32422428305624\\
2.5	1.36170647625638\\
2.75	1.39224103953059\\
3	1.41641241447443\\
3.25	1.43613343234414\\
3.5	1.4521470310643\\
3.75	1.46564804875777\\
4	1.47716304017593\\
4.25	1.48666135653807\\
4.5	1.49495135543784\\
4.75	1.50165791867292\\
5	1.50807826263434\\
5.25	1.51286119146033\\
5.5	1.5178495862579\\
5.75	1.52201189244416\\
6	1.52582519950535\\
6.25	1.52814197548412\\
6.5	1.53182947490725\\
6.75	1.53338424984065\\
7	1.53595496035007\\
7.25	1.53809345693238\\
7.5	1.54033509093984\\
7.75	1.54185860670808\\
8	1.54363139995616\\
8.25	1.54496160421793\\
8.5	1.5457712772443\\
8.75	1.54702311590485\\
9	1.54857803322811\\
9.25	1.54935543966834\\
9.5	1.55074343593564\\
9.75	1.55104451865318\\
10	1.55164393366289\\
20	1.56313959345703\\
30	1.56596178120805\\
40	1.56587212819918\\
50	1.56673875264517\\
};
\addlegendentry{ $\mmse(\vec{\mathbf{X}}|\vec{\mathbf{Y}})$}

\addplot [color=black, dashdotted, line width=2.0pt]
  table[row sep=crcr]{%
0	-1.77635683940025e-14\\
0.25	0.206588649711154\\
0.5	0.756730202870888\\
0.75	1.35079987244675\\
1	1.77851305564864\\
1.25	2.05039813023699\\
1.5	2.23450637565206\\
1.75	2.36999732680382\\
2	2.47476734178792\\
2.25	2.5571186270796\\
2.5	2.62314248945869\\
2.75	2.67602579864787\\
3	2.71969330396518\\
3.25	2.75538756757577\\
3.5	2.78485871140489\\
3.75	2.80954972143298\\
4	2.83029956168669\\
4.25	2.84790698514079\\
4.5	2.86316866902377\\
4.75	2.87621490992933\\
5	2.88751527911859\\
5.25	2.89743862580337\\
5.5	2.90605309978259\\
5.75	2.91362716537599\\
6	2.92021741040752\\
6.25	2.92631324936662\\
6.5	2.93174052535131\\
6.75	2.93650135428711\\
7	2.94061575002615\\
7.25	2.9448280384312\\
7.5	2.94809935796426\\
7.75	2.95136578335889\\
8	2.95426525383283\\
8.25	2.95685410782557\\
8.5	2.9594496417988\\
8.75	2.96185723035813\\
9	2.96373930793813\\
9.25	2.96575872731077\\
9.5	2.96729551104164\\
9.75	2.96907845180199\\
10	2.97040036148795\\
20	2.99236217652746\\
30	2.99675667639819\\
40	2.99807555982224\\
50	2.99857887499217\\
};
\addlegendentry{\text{MSE} of ${\bf  \hat{f}}$}

\addplot [color=black, dotted,  line width=2.0pt]
  table[row sep=crcr]{%
0	1.5676\\
0.25	1.5676\\
0.5	1.5676\\
0.75	1.5676\\
1	1.5676\\
1.25	1.5676\\
1.5	1.5676\\
5	1.5676\\
5.25	1.5676\\
7	1.5676\\
7.25	1.5676\\
8.25	1.5676\\
9.25	1.5676\\
9.75	1.5676\\
10	1.5676\\
20	1.5676\\
30	1.5676\\
40	1.5676\\
50	1.5676\\
};
\addlegendentry{$\mathsf{Var}(\vec{\mathbf{X}})$ or MSE of $ {\bf  \hat{h}}$}

\addplot [color=black, dashed,  line width=2.0pt]
  table[row sep=crcr]{%
0.2	0.138823551577924\\
0.25	0.204089399807681\\
0.3	0.276600021434491\\
0.35	0.378261513067393\\
0.4	0.485474475118409\\
0.45	0.611011861426645\\
0.5	0.750807157184475\\
0.55	0.887410157753891\\
0.6	1.0555246789322\\
0.65	1.22670569740734\\
0.7	1.40010162178783\\
0.75	1.59166802121692\\
0.8	1.79433461131592\\
0.85	1.97335668377047\\
0.9	2.20129101124653\\
0.95	2.40422690690099\\
1	2.65207406899572\\
1.05	2.85408811947552\\
1.1	3.10799427249461\\
1.15	3.37150933326584\\
1.2	3.64714415826815\\
1.25	3.81784714494871\\
1.3	4.16042789949964\\
1.35	4.4244074689597\\
1.4	4.70806172623998\\
1.45	4.99118587870554\\
1.5	5.30967181247786\\
};
\addlegendentry{MSE of MLE}

\end{axis}  
\end{tikzpicture}%

\caption{$n=3$}
\vspace{0.8cm}
\end{subfigure}
\hfill

\begin{subfigure}[t]{0.45\textwidth}
%
%
\pgfplotsset{every axis plot/.append style={very thick}}

\definecolor{mycolor1}{rgb}{0.00000,0.45000,0.74000}%

\begin{tikzpicture}

\begin{axis}[%
width=9cm,
height=6.5cm,
at={(1.73in,1.152in)},
xmin=0,
xmax=10,
xlabel style={font=\color{white!15!black},below=-2mm},
xlabel={$\sigma$},
ymin=0,
ymax=4.5,
ylabel style={font=\color{white!15!black}},
ylabel style={at={(axis description cs:0.09,.5)},anchor=south},
ylabel={MSE},
axis background/.style={fill=white},
xmajorgrids,
ymajorgrids,
legend style={at={(0.98,0.8)}, align=left, draw=white!15!black, font=\footnotesize}
]

\addplot [color=black, line width=2.0pt]
  table[row sep=crcr]{%
0	1.17239551400417e-13\\
0.25	0.188300809178018\\
0.5	0.522290677790747\\
0.75	0.798198472913008\\
1	0.992225190782055\\
1.25	1.13449397879503\\
1.5	1.24239609633851\\
1.75	1.32550090737312\\
2	1.39054980560255\\
2.25	1.4421447718515\\
2.5	1.48173584374556\\
2.75	1.51475861080486\\
3	1.5405070864252\\
3.25	1.56140518509445\\
3.5	1.58040814481244\\
3.75	1.59368815265671\\
4	1.60632551802694\\
4.25	1.61654684990109\\
4.5	1.62468344634815\\
4.75	1.63283938893661\\
5	1.63961192351537\\
5.25	1.64504219019781\\
5.5	1.64978868847845\\
5.75	1.65400710131697\\
6	1.65765224201517\\
6.25	1.66187861035722\\
6.5	1.6646049830277\\
6.75	1.66795605491003\\
7	1.67006777119367\\
7.25	1.67297799063884\\
7.5	1.67453575863979\\
7.75	1.67709725330362\\
8	1.67789142000456\\
8.25	1.67956626962249\\
8.5	1.68023776375483\\
8.75	1.68260256298087\\
9	1.68318846636139\\
9.25	1.68408519334614\\
9.5	1.68536025926973\\
9.75	1.68540531833664\\
10	1.68642147296434\\
20	1.69974001195765\\
30	1.70188716553296\\
40	1.70315776455397\\
50	1.70302769799307\\
};
\addlegendentry{$\mmse(\vec{\mathbf{X}}|\vec{\mathbf{Y}})$}

\addplot [color=black, dashdotted, line width=2.0pt]
  table[row sep=crcr]{%
0	-2.48689957516035e-14\\
0.25	0.291398375777379\\
0.5	1.07533537350432\\
0.75	1.88879205701736\\
1	2.44071646026927\\
1.25	2.780527209768\\
1.5	3.00956598638681\\
1.75	3.17968519039129\\
2	3.31249359455842\\
2.25	3.41845084685221\\
2.5	3.50368403068138\\
2.75	3.57288180451444\\
3	3.62948989659117\\
3.25	3.67632317191734\\
3.5	3.71508537899995\\
3.75	3.74753276492221\\
4	3.77492012056351\\
4.25	3.79831894801755\\
4.5	3.81839145183463\\
4.75	3.83559795475836\\
5	3.85051470224314\\
5.25	3.86364181275512\\
5.5	3.87506114054235\\
5.75	3.8850350830437\\
6	3.89406467277594\\
6.25	3.90206849031941\\
6.5	3.90918046678383\\
6.75	3.91543359582385\\
7	3.92117078303182\\
7.25	3.92633678399244\\
7.5	3.93106893839188\\
7.75	3.93540465422776\\
8	3.93912206559076\\
8.25	3.94264068152487\\
8.5	3.9459965494429\\
8.75	3.94885109300179\\
9	3.9516583264068\\
9.25	3.95421444219969\\
9.5	3.95661153626715\\
9.75	3.95867232982404\\
10	3.96070200975747\\
20	3.99001460378017\\
30	3.99557237438405\\
40	3.99751610669426\\
50	3.99840288220112\\
};

\addlegendentry{$\text{MSE}$ of ${\bf  \hat{f}}$}

\addplot [color=black, dotted,  line width=2.0pt]
  table[row sep=crcr]{%
0	1.704\\
0.25	1.704\\
0.5	1.704\\
0.75	1.704\\
1	1.704\\
1.25	1.704\\
1.5	1.704\\
5	1.704\\
5.25	1.704\\
7	1.704\\
7.25	1.704\\
8.25	1.704\\
9.25	1.704\\
9.75	1.704\\
10	1.704\\
20	1.704\\
30	1.704\\
40	1.704\\
50	1.704\\
};
\addlegendentry{$\mathsf{Var}(\vec{\mathbf{X}})$ or MSE of $ {\bf  \hat{h}}$}

\addplot [color=black, dashed,  line width=2.0pt ]
  table[row sep=crcr]{%
0.2	0.236926675152555\\
0.25	0.297267630086909\\
0.3	0.378407657972022\\
0.35	0.501607321376547\\
0.4	0.655855485333737\\
0.45	0.81359434452868\\
0.5	0.994599008212105\\
0.55	1.18094347028836\\
0.6	1.3820489089996\\
0.65	1.58728485367763\\
0.7	1.83193143446518\\
0.75	2.06015935560811\\
0.8	2.30782031512332\\
0.85	2.55952747070104\\
0.9	2.8040403606818\\
0.95	3.07291627037593\\
1	3.36273176552742\\
1.05	3.64195065270512\\
1.1	3.90396668398953\\
1.15	4.25702284559366\\
1.2	4.56296120214\\
1.25	4.80471568242277\\
1.3	5.20009955345031\\
1.35	5.47156613474357\\
1.4	5.83027727780403\\
1.45	6.25770090036599\\
1.5	6.59315640701413\\
1.55	6.9290512976315\\
1.6	7.28204342410917\\
1.65	7.7491878460711\\
1.7	8.10097630055143\\
};
\addlegendentry{MSE of MLE}

\end{axis}  
\end{tikzpicture}%
\caption{$n=4$}
\end{subfigure}\quad \quad
\begin{subfigure}[t]{0.45\textwidth}
%
%

\pgfplotsset{every axis plot/.append style={very thick}}

\begin{tikzpicture}

\begin{axis}[%
width=9cm,
height=6.5cm,
at={(1.73in,1.152in)},
xmin=0,
xmax=30,
xlabel style={font=\color{white!15!black},below=-2mm},
xlabel={$n$},
ymin=0,
ymax=1.2,
ylabel style={font=\color{white!15!black}, at={(0.05,0.5)}},  
ylabel={ $ \frac{\mathsf{Var}(\vec{\mathbf{X}})}{n}$},
axis background/.style={fill=white},
xmajorgrids,
ymajorgrids
]
\addplot [color=black, forget plot, line width=2.0pt]
  table[row sep=crcr]{%
1	1.00272553858683\\
2	0.681795682004736\\
3	0.521732500961701\\
4	0.426948090000691\\
5	0.360913292112365\\
6	0.313768126223787\\
7	0.278132071989545\\
8	0.249912848632862\\
9	0.227711211090029\\
10	0.208560550134954\\
11	0.192701354524409\\
12	0.179382244836214\\
13	0.167708788033509\\
14	0.157378473320193\\
15	0.148492902630509\\
16	0.140647965078257\\
17	0.13352654563728\\
18	0.127045202251083\\
19	0.121382763490684\\
20	0.116049019741956\\
21	0.111255903378043\\
22	0.106758962592807\\
23	0.102856831587748\\
24	0.0990815171158099\\
25	0.0955913155448726\\
26	0.092322350509708\\
27	0.0894208284115213\\
28	0.0866546261260253\\
29	0.0839581129149711\\
30	0.0815263809881803\\
};
\end{axis}
\end{tikzpicture}%
\caption{Variance of $\vec{\mathbf{X}}$ normalized by $n$ vs. $n$. }
\label{fig:Var}
\end{subfigure}

\caption{Comparison of the $\mmse(\vec{\mathbf{X}}|\vec{\mathbf{Y}})$ and the MSE of ${\bf  \hat{h}}$, ${\bf  \hat{f}}$  and $  {\bf  \hat{\boldsymbol{\ell}}}_{\text{MLE}}$  versus $\sigma$, and comparison of   $\mathsf{Var}(\vec{\mathbf{X}})/n$ versus $n$.  }
\label{fig:MMSEcurves}
\vspace{-0.3cm}
\end{figure*}
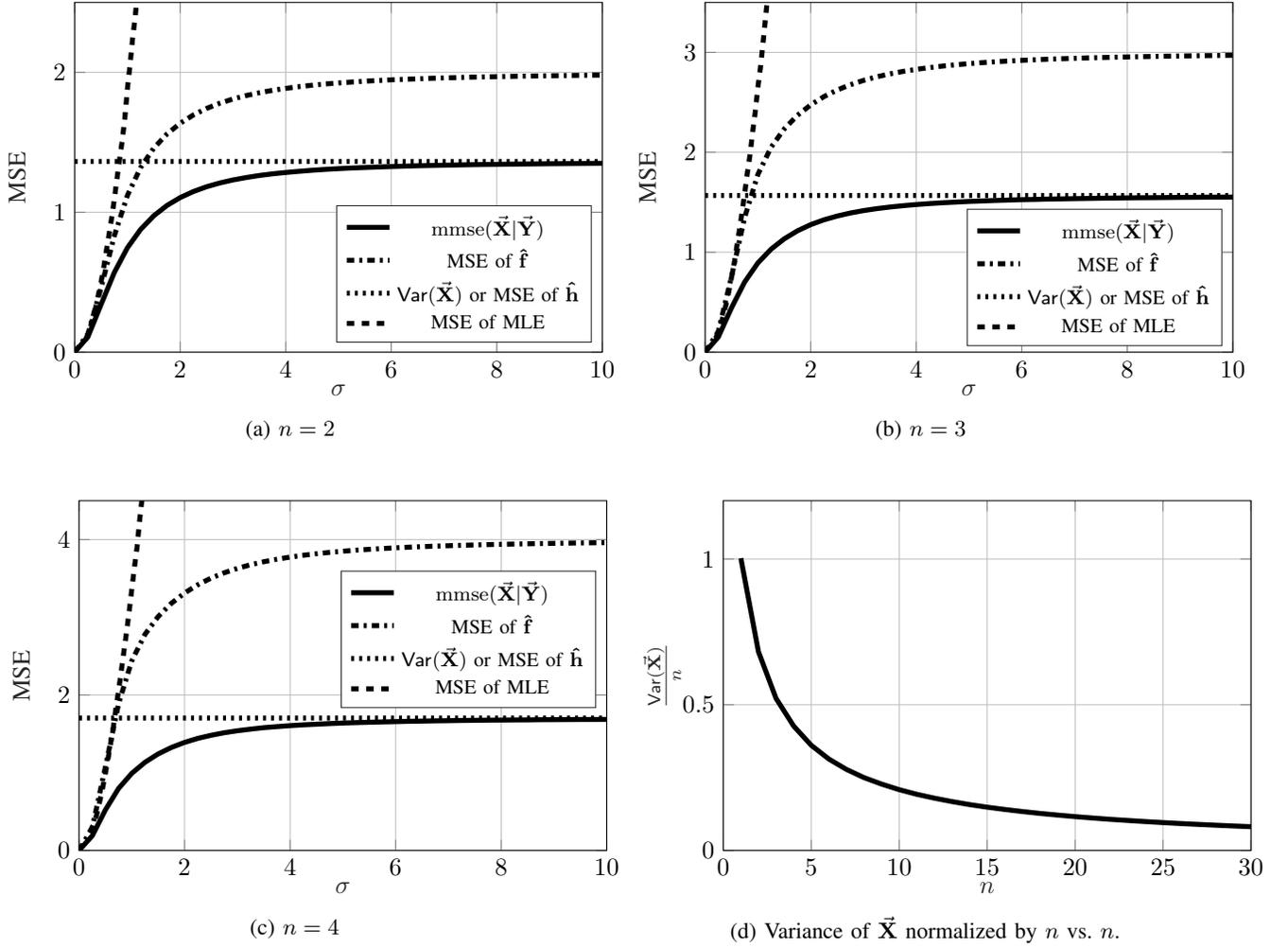

 We start by noting that, under the assumption that $\X \sim \mathcal{N}({\bf 0}_n,   \mathbf{I}_n)$, the proposed suboptimal estimator in~\eqref{eq:ProposeSubOptimalEstimator} becomes
\begin{equation}
\hat{f}_k( \vec{\mathbf{y}})=  \sum_{ \pi \in  \mathcal{P} } a_{\pi}   \left(\vec{\mathbf{y}}\right)     \left[\mathbf{P}_{\pi}\vec{\mathbf{y}} \right]_k  ,    \label{eq:linearLike}
\end{equation}
where $\left[\mathbf{P}_{\pi}\vec{\mathbf{y}} \right]_k  $ is $k$-th component of the vector $\mathbf{P}_{\pi}\vec{\mathbf{y}}$ and where
\begin{equation}
a_{\pi}   \left(\vec{\mathbf{y}}\right)  =   \frac{\sigma}{1+\sigma^2}  \mathbb{P} \left[  \mathbf{X} \in \vec{\mathbb{R}}_n |  \mathbf{Y}=\mathbf{P}_{\pi}\vec{\mathbf{y}}      \right] . 
\end{equation}
Fig.~\ref{fig:MMSEcurves} compares the performance, in terms of the MSE, of the suboptimal estimator in~\eqref{eq:linearLike} (dashed-dotted curve)  and the optimal estimator in~\eqref{eq:EstimatorBothXandYsymmetric} (solid curve), versus different values of $\sigma$. From Fig.~\ref{fig:MMSEcurves} we observe that the suboptimal estimator  in~\eqref{eq:linearLike}  performs closely to the optimal estimator in~\eqref{eq:EstimatorBothXandYsymmetric} for small values of $\sigma$. However, for higher values of $\sigma$, Fig.~\ref{fig:MMSEcurves} suggests that the MSE of the suboptimal estimator in~\eqref{eq:linearLike} is to within a constant gap of the MSE of the optimal estimator.
We next formalize these observations in Theorem~\ref{thm:ErrorInGaussianNoiseCase}.



\begin{thm}\label{thm:ErrorInGaussianNoiseCase} Let $\X$ be exchangeable and $\Y| \X=\mathbf{x} \sim \mathcal{N}({\bf x}, \sigma^2  \mathbf{I}_n)$. Assume that $\E[X_k^2]<\infty$ for all $k \in [n]$.  Then, the approximation error  in~\eqref{eq:GapToOptimalityOfSubOptmalEstimator} satisfies the following
\begin{subequations}
\begin{align}
\lim_{ \sigma \to 0}  \Delta_{\text{up}} &=0, \label{DeltaSigmaToZero}\\
\lim_{ \sigma \to \infty }  \Delta_{\text{up}} &=   \E \left[ \|\X \|^2 \right] \left (1- \frac{1}{n!} \right ). \label{DeltaSigmaToInf} 
\end{align}
\end{subequations}
\end{thm}
\begin{IEEEproof}
We start by noting that, by a simple application of the dominated convergence theorem, we have that
\begin{align}
\lim_{ \sigma \to 0 } \mathbb{P} \left[  \mathbf{X} \in \vec{\mathbb{R}}_n |  \mathbf{Y}=\mathbf{y}    \right] &=    1_{\vec{\mathbb{R}}_n}(\mathbf{y}) \label{eq:ConditionaDistSigmaZero},  \\
\lim_{ \sigma \to \infty } \mathbb{P} \left[  \mathbf{X} \in \vec{\mathbb{R}}_n |  \mathbf{Y}=\mathbf{y}    \right] &=   \mathbb{P} \left[  \mathbf{X} \in \vec{\mathbb{R}}_n   \right]= \frac{1}{n!}. \label{eq:ConditionaDistSigmaInfinity}
\end{align} 

Now observe that
\begin{align}
 &\lim_{ \sigma \to 0 }   \mathbb{E} \left [  X_k^2 \cdot g \left( \mathbf{P}_{\pi}\mathbf{Y}  \right)     \mid    \mathbf{Y} \in \vec{\mathbb{R}}_n   \right ]  \notag\\
 =& n! \ \lim_{ \sigma \to 0 }   \mathbb{E} \left [  X_k^2 \cdot g \left( \mathbf{P}_{\pi}\mathbf{Y}  \right)      \cdot  1_{\vec{\mathbb{R}}_n}(\Y)  \right ]  \notag \\
 \stackrel{{\rm{(a)}}}{=}& n! \ \mathbb{E} \left [ \lim_{ \sigma \to 0 }    X_k^2 \cdot g \left( \mathbf{P}_{\pi}\mathbf{Y}  \right)     \cdot  1_{\vec{\mathbb{R}}_n}(\Y)  \right ]\notag   \\
  \stackrel{{\rm{(b)}}}{=}& n! \ \mathbb{E} \left [    X_k^2 \cdot 1_{\vec{\mathbb{R}}_n} \left(\mathbf{P}_{\pi} \mathbf{X} \right)  \left(   1   - 1_{\vec{\mathbb{R}}_n} \left(\mathbf{P}_{\pi} \mathbf{X} \right)    \right)   \cdot  1_{\vec{\mathbb{R}}_n}(\mathbf{X})  \right ]  \notag  \\
  \stackrel{{\rm{(c)}}}{=} & 0, 
\label{eq:LimitsOnGapWhenSigmaZero}
 \end{align}
 where the labeled equalities follow from: ${\rm{(a)}}$ using the dominated convergence theorem with the bound 
 \begin{align}
 &X_k^2 \cdot g\left( \mathbf{P}_{\pi}\mathbf{Y}  \right) \cdot  1_{\vec{\mathbb{R}}_n}(\Y)  \notag\\
 =\! & X_k^2 \cdot p_{\vec{\mathbf{X}}|  \mathbf{Y}} \left( \mathbf{P}_{\pi}\mathbf{Y}  \right)   \left(   1   \!-\! p_{\vec{\mathbf{X}}|  \mathbf{Y}}( \mathbf{P}_{\pi}\mathbf{Y}  )   \right)   \cdot  1_{\vec{\mathbb{R}}_n}(\Y)  
 \!\le\!   X_k^2, \label{eq:BoundForDCT}
 \end{align} 
with
$
p_{\vec{\mathbf{X}}|  \mathbf{Y}}( \mathbf{P}_{\pi}\vec{\mathbf{y}} )= \mathbb{P} \left[  \mathbf{X} \in \vec{\mathbb{R}}_n |  \mathbf{Y}=\mathbf{P}_{\pi}\vec{\mathbf{y}}      \right],
$
and since $\E[X_k^2] < \infty$; ${\rm{(b)}}$ using the limit in \eqref{eq:ConditionaDistSigmaZero}; and ${\rm{(c)}}$ using the fact that $1_{\vec{\mathbb{R}}_n} \left(\mathbf{P}_{\pi} \mathbf{X} \right) \left(   1   - 1_{\vec{\mathbb{R}}_n} \left(\mathbf{P}_{\pi} \mathbf{X} \right)    \right)   =0 $.
Combining~\eqref{eq:LimitsOnGapWhenSigmaZero} with the definition of $\Delta_{\text{up}}$ in~\eqref{eq:GapToOptimalityOfSubOptmalEstimator} we obtain~\eqref{DeltaSigmaToZero}.

We now focus on the case of $ \sigma \to \infty $, and we obtain
 \begin{align}
 &\lim_{ \sigma \to \infty }   \mathbb{E} \left [  X_k^2 \cdot g \left( \mathbf{P}_{\pi}\mathbf{Y}  \right)      \mid    \mathbf{Y} \in \vec{\mathbb{R}}_n   \right ]  \notag\\
 \stackrel{{\rm{(a)}}}{=} &   \lim_{ \sigma \to \infty }  \mathbb{E} \left [     X_k^2 \cdot g \left( \mathbf{P}_{\pi} \vec{ \mathbf{Y}} \right)        \right ] \notag \\
 \stackrel{{\rm{(b)}}}{=} &  \mathbb{E} \left [     \lim_{ \sigma \to \infty }   X_k^2 \cdot g \left( \mathbf{P}_{\pi} \vec{ \mathbf{Y}} \right)        \right ] \notag \\
  \stackrel{{\rm{(c)}}}{=} &   \mathbb{E} \left [     X_k^2 \cdot  \frac{1}{n!}  \left(   1  -\frac{1}{n!}   \right)     \right ] \notag \\
  = &   \mathbb{E} \left [     X_k^2     \right ]  \cdot  \frac{1}{n!}  \left(   1  -\frac{1}{n!}   \right), \label{eq:LimitsOnGapWhenSigmaInfinity}
 \end{align} 
where the labeled equalities follow from: ${\rm{(a)}}$ the fact that when $\sigma \to \infty$ $\X$ and $\Y$ are independent; ${\rm{(b)}}$ using the dominated convergence theorem with the bound in~\eqref{eq:BoundForDCT} and the assumption that $\E[X_k^2] < \infty$; and ${\rm{(c)}}$ using the limit in~\eqref{eq:ConditionaDistSigmaInfinity}.
Combining~\eqref{eq:LimitsOnGapWhenSigmaInfinity} with $\Delta_{\text{up}}$ in~\eqref{eq:GapToOptimalityOfSubOptmalEstimator}, we obtain~\eqref{DeltaSigmaToInf} .

This concludes the proof of Theorem~\ref{thm:ErrorInGaussianNoiseCase}.
\end{IEEEproof}

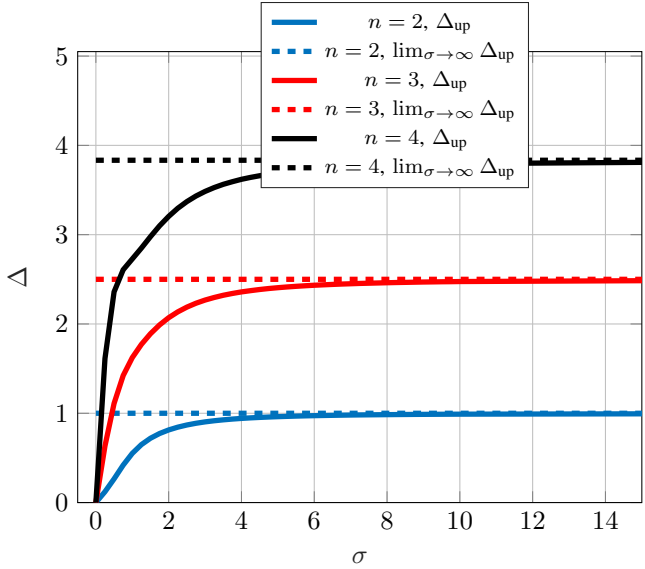
\begin{figure}[h!]
\center
%
%
\definecolor{mycolor1}{rgb}{0.00000,0.44706,0.74118}%
\definecolor{mycolor2}{rgb}{0.00000,0.44700,0.74100}%

\begin{tikzpicture}

\begin{axis}[%
width=7.5cm,
height=6cm,
at={(1.73in,1.152in)},
scale only axis,
xmin=-0.5,
xmax=15,
xlabel style={font=\color{white!15!black}},
xlabel={$\sigma$},
ymin=0,
ymax=5.05,
ylabel style={font=\color{white!15!black}},
ylabel style={at={(axis description cs:0.09,.5)},anchor=south},
ylabel={$\Delta$},
axis background/.style={fill=white},
xmajorgrids,
ymajorgrids,
legend style={at={(0.8,1.11)}, align=left, draw=white!15!black, font=\footnotesize}
]
\addplot [color=mycolor1, line width=2.0pt]
  table[row sep=crcr]{%
0	0\\
0.25	0.121717108642228\\
0.5	0.265311572740242\\
0.75	0.419301136906241\\
1	0.550800860084417\\
1.25	0.648888428126124\\
1.5	0.72032593373884\\
1.75	0.773263904905766\\
2	0.813363610829702\\
2.25	0.844373425612544\\
2.5	0.86870595115248\\
2.75	0.887954908555432\\
3	0.903591473424076\\
3.25	0.916175370277547\\
3.5	0.92646183624155\\
3.75	0.935059820414282\\
4	0.942388666807115\\
4.25	0.94851431174328\\
4.5	0.953570972362131\\
4.75	0.95806114148506\\
5	0.961926641917447\\
5.25	0.965235983002178\\
5.5	0.968305670835419\\
5.75	0.970911235757375\\
6	0.97320548545514\\
6.25	0.975135861397502\\
6.5	0.976967090284996\\
6.75	0.978646435415151\\
7	0.98018007217798\\
7.25	0.981456920892373\\
7.5	0.98252808528426\\
7.75	0.983496975167713\\
8	0.984574775879822\\
8.25	0.985590855422208\\
8.5	0.986378924299781\\
8.75	0.987233683379057\\
9	0.987781835504326\\
9.25	0.988415293327515\\
9.5	0.988957463148593\\
9.75	0.989466530198493\\
10	0.989940258598557\\
20	0.997510632839544\\
30	0.998791840292648\\
40	0.999271132697773\\
50	0.999515082806989\\
};
\addlegendentry{$n=2$, $\Delta_{\text{up}}$}

\addplot [color=mycolor2, dashed, line width=2.0pt]
  table[row sep=crcr]{%
0	1\\
5.55555555555556	1\\
11.1111111111111	1\\
16.6666666666667	1\\
22.2222222222222	1\\
27.7777777777778	1\\
33.3333333333333	1\\
38.8888888888889	1\\
44.4444444444444	1\\
50	1\\
};
\addlegendentry{$n=2$, $\lim_{\sigma \to \infty} \Delta_{\text{up}}$}

\addplot [color=red, line width=2.0pt]
  table[row sep=crcr]{%
0	0\\
0.25	0.634037035296269\\
0.5	1.11026137895512\\
0.75	1.42220282987447\\
1	1.62298112932762\\
1.25	1.77098165674407\\
1.5	1.89077020571516\\
1.75	1.98926494401547\\
2	2.07012769635613\\
2.25	2.1354667227817\\
2.5	2.1886826970353\\
2.75	2.2322288691152\\
3	2.26775725436254\\
3.25	2.29658125371902\\
3.5	2.32149471525481\\
3.75	2.34163490899513\\
4	2.35911334712184\\
4.25	2.37309845579135\\
4.5	2.38602675578304\\
4.75	2.39706589881971\\
5	2.40624107624041\\
5.25	2.41445783898364\\
5.5	2.42152237357797\\
5.75	2.42790853361685\\
6	2.4336516882472\\
6.25	2.43830821336649\\
6.5	2.44281211897629\\
6.75	2.44692197011143\\
7	2.45032486304658\\
7.25	2.45393783199552\\
7.5	2.45643222108604\\
7.75	2.45908442925128\\
8	2.46172743374529\\
8.25	2.46408529809207\\
8.5	2.46580501389585\\
8.75	2.46779837133089\\
9	2.4697458377557\\
9.25	2.47108328901035\\
9.5	2.47264066588251\\
9.75	2.4735479078399\\
10	2.47516456765143\\
20	2.49389531731811\\
30	2.4970992676261\\
40	2.49849268564044\\
50	2.49899315042159\\
};
\addlegendentry{$n=3$, $\Delta_{\text{up}}$ }

\addplot [color=red, dashed, line width=2.0pt]
  table[row sep=crcr]{%
0	2.5\\
5.55555555555556	2.5\\
11.1111111111111	2.5\\
16.6666666666667	2.5\\
22.2222222222222	2.5\\
27.7777777777778	2.5\\
33.3333333333333	2.5\\
38.8888888888889	2.5\\
44.4444444444444	2.5\\
50	2.5\\
};
\addlegendentry{$n=3$, $\lim_{\sigma \to \infty} \Delta_{\text{up}}$}

\addplot [color=black, line width=2.0pt]
  table[row sep=crcr]{%
0	0\\
0.25	1.61349837236954\\
0.5	2.35950132721835\\
0.75	2.60693126402373\\
1	2.72798095531406\\
1.25	2.84948730419594\\
1.5	2.97795783914671\\
1.75	3.09909462783851\\
2	3.20471996546945\\
2.25	3.29557413220717\\
2.5	3.3704193642022\\
2.75	3.43283213006135\\
3	3.48428713725166\\
3.25	3.52745107996392\\
3.5	3.56381940862998\\
3.75	3.59324135995519\\
4	3.61998223751859\\
4.25	3.64141626872759\\
4.5	3.65998541458174\\
4.75	3.67660002673807\\
5	3.69112061064657\\
5.25	3.70257247897086\\
5.5	3.71404388204223\\
5.75	3.72291800018732\\
6	3.73199732694469\\
6.25	3.73855202854092\\
6.5	3.74614160958141\\
6.75	3.75216816739165\\
7	3.75779625003882\\
7.25	3.76266442914728\\
7.5	3.76763971593683\\
7.75	3.77111551759001\\
8	3.77438059853761\\
8.25	3.77861876368242\\
8.5	3.78121990140113\\
8.75	3.78412447801239\\
9	3.78672781273783\\
9.25	3.78892528313882\\
9.5	3.79105441880154\\
9.75	3.79323301688162\\
10	3.79554560461326\\
20	3.82396079465864\\
30	3.82867215949973\\
40	3.83000308302914\\
50	3.83148545419444\\
};
\addlegendentry{ $n=4$, $\Delta_{\text{up}}$ }

\addplot [color=black, dashed, line width=2.0pt]
  table[row sep=crcr]{%
0	3.83333333333333\\
5.55555555555556	3.83333333333333\\
11.1111111111111	3.83333333333333\\
16.6666666666667	3.83333333333333\\
22.2222222222222	3.83333333333333\\
27.7777777777778	3.83333333333333\\
33.3333333333333	3.83333333333333\\
38.8888888888889	3.83333333333333\\
44.4444444444444	3.83333333333333\\
50	3.83333333333333\\
};
\addlegendentry{$n=4$, $\lim_{\sigma \to \infty} \Delta_{\text{up}}$}

\end{axis}
\end{tikzpicture}%
\caption{Plot of $\Delta_{\text{up}}$ evaluated with $\X \sim \mathcal{N}( {\bf 0}_n,  {\bf I}_n)$  versus $\sigma$. }
\label{fig:Deltas}
\vspace{-0.3cm}
\end{figure}

 As highlighted above, Theorem~\ref{thm:ErrorInGaussianNoiseCase}
shows that the suboptimal estimator in~\eqref{eq:linearLike} is asymptotically optimal in the low noise regime,  and  its MSE performance is always to within a constant gap of the optimal MMSE.
The plot of the upper bound on the penalty in~\eqref{eq:GapToOptimalityOfSubOptmalEstimator} for the Gaussian input $\X\sim \mathcal{N}( {\bf 0}_n,  {\bf I}_n)$ is shown in Fig.~\ref{fig:Deltas},  versus different values of~$\sigma$.

We now turn our attention to analyzing the performance of the suboptimal estimator in~\eqref{eq:EquivalentHighNoise}. In particular, we start by showing that this estimator is optimal in the high noise regime, as stated in the following theorem.
\begin{thm} 
\label{thm:EstPerHighNoise}
Let $\X$ be exchangeable and $\Y| \X=\mathbf{x} \sim \mathcal{N}({\bf x}, \sigma^2  \mathbf{I}_n)$. Assume that $\E[X_k^2]<\infty$ for all $k \in [n]$.  Then, the approximation error  in~\eqref{eq:ErrorHighSnr} satisfies the following:
\begin{align}
\label{eq:LimSigma}
\lim_{\sigma \to \infty}  \Delta_\sigma=0. 
\end{align}  
\end{thm}
\begin{IEEEproof}
To show~\eqref{eq:LimSigma}, we have to verify that
the limit in~\eqref{eq:ConitinoutyAssumption} holds. 
For fixed values of $k$  and ${\bf y}$, we have
\begin{align*}
\lim_{\sigma \to \infty}  \E[ X_k |\Y={\bf y} ]& \stackrel{\rm (a)}{=}   \lim_{\sigma \to \infty}  \E[ X_k |Y_k=y_k] \notag\\
&\stackrel{\rm (b)}{=}  \lim_{\sigma \to \infty}  \frac{ \E \left [    X_k \frac{1}{\sqrt{ 2 \pi \sigma^2}} \eu^{- \frac{(y-X_k)^2}{2 \sigma^2}  }  \right]}{ \E \left [     \frac{1}{\sqrt{ 2 \pi \sigma^2}} \eu^{- \frac{(y-X_k)^2}{2 \sigma^2}  }  \right] }\\
&\stackrel{\rm (c)}{=}  \!  \frac{ \E \left [    X_k  \lim_{\sigma \to \infty}  \eu^{- \frac{(y-X_k)^2}{2 \sigma^2}  }  \right]}{ \E \left [   \lim_{\sigma \to \infty}   \eu^{- \frac{(y-X_k)^2}{2 \sigma^2}  }  \right] }\!= \!   \frac{ \E \left [    X_k   \right]}{ 1 },
\end{align*} 
where the labeled equalities follow from: ${\rm (a)} $ the fact that $\Y$ only depends on $X_k$ through $Y_k$; ${\rm (b)} $  using Bayes rule;  and ${\rm (c)} $ using $\eu^{- \frac{(y-X_k)^2}{2 \sigma^2}} \le 1$ to apply  the dominated convergence theorem.
This concludes the  proof of Theorem~\ref{thm:EstPerHighNoise}. 
\end{IEEEproof}
In Fig.~\ref{fig:MMSEcurves}, we compare the performance of the suboptimal estimator in~\eqref{eq:EquivalentHighNoise} (dotted curve)  and the optimal one in~\eqref{eq:EstimatorBothXandYsymmetric} (solid curve), versus different values of $\sigma$. 
From Fig.~\ref{fig:MMSEcurves} we observe that, as also proved in Theorem~\ref{thm:EstPerHighNoise}, the suboptimal estimator  in~\eqref{eq:EquivalentHighNoise}  performs closely to the optimal estimator in~\eqref{eq:EstimatorBothXandYsymmetric} for high values of $\sigma$. We also highlight that, as shown in~\eqref{eq:MSESubHighNoise}, the MSE of the suboptimal estimator in~\eqref{eq:EquivalentHighNoise} is simply the variance of $\vec{\X}$, i.e., it is a constant with respect to $\sigma$. 

We conclude this section by analyzing the MLE in~\eqref{eq:MaximumLikelihood_Estimator}, whose structure for the Gaussian noise channel is given next.
\begin{lem} 
\label{lem:MLEGaussi}
Let $\Y| \X=\mathbf{x} \sim \mathcal{N}({\bf x}, \sigma^2  \mathbf{I}_n)$.  Then, the MLE is given by 
\begin{align*}
 {\bf  \hat{\boldsymbol{\ell}}}_{\text{MLE}}( \vec{\mathbf{y}})  \in  \left \{  \mathbf{t} \in  \vec{\mathbb{R}}_n:  \sum_{ \pi \in  \mathcal{P} } \mathbf{P}_{\pi}\vec{\mathbf{y}} \eu^{  \frac{(\mathbf{P}_{\pi}\vec{\mathbf{y}})^T  \mathbf{t}}{\sigma^2} } =  \mathbf{t}  \sum_{ \pi \in  \mathcal{P} }  \eu^{ \frac{ (\mathbf{P}_{\pi}\vec{\mathbf{y}})^T  \mathbf{t} }{\sigma^2} }   \right \}.  
\end{align*} 
If the set above is empty, then  set  $  {\bf  \hat{\boldsymbol{\ell}}}_{\text{MLE}}( \vec{\mathbf{y}}) =\vec{\mathbf{y}}$. 
\end{lem} 
\begin{IEEEproof}
Specializing~\eqref{eq:MaximumLikelihood_Estimator} to the Gaussian noise channel, a maximum in~\eqref{eq:MaximumLikelihood_Estimator} must satisfy the following equation
\begin{equation*}
\mathbf{0}_n=  \nabla_{ \mathbf{t}} \sum_{ \pi \in  \mathcal{P} }  f_{\mathbf{Y}|\mathbf{X} } (\mathbf{P}_{\pi}\vec{\mathbf{y}} | \mathbf{t})\!=\!  \sum_{ \pi \in  \mathcal{P} }  \frac{(\mathbf{P}_{\pi}\vec{\mathbf{y}} - \mathbf{t})}{ \sigma^2} f_{\mathbf{Y}|\mathbf{X} } (\mathbf{P}_{\pi}\vec{\mathbf{y}} | \mathbf{t}),
\end{equation*}
for $\mathbf{t} \in  \vec{\mathbb{R}}_n$.
The expression above is equivalent to 
\begin{equation}
 \sum_{ \pi \in  \mathcal{P} } \mathbf{P}_{\pi}\vec{\mathbf{y}}  f_{\mathbf{Y}|\mathbf{X} } (\mathbf{P}_{\pi}\vec{\mathbf{y}} | \mathbf{t})=  \mathbf{t}  \sum_{ \pi \in  \mathcal{P} } f_{\mathbf{Y}|\mathbf{X} } (\mathbf{P}_{\pi}\vec{\mathbf{y}} | \mathbf{t}). 
\end{equation} 
The proof is concluded by observing that $\| \mathbf{P}_{\pi}\vec{\mathbf{y}}\| =\|\mathbf{y}\|$,
\begin{align*}
(2 \pi \sigma^2)^{ \frac{n}{2} }f_{\mathbf{Y}|\mathbf{X} } (\mathbf{P}_{\pi}\vec{\mathbf{y}} | \mathbf{t})\!=\!  \eu^{-\frac{\| \mathbf{P}_{\pi}\vec{\mathbf{y}} - \mathbf{t}\|^2 }{2 \sigma^2}}\!=\!\eu^{-\frac{\|\mathbf{y}\|^2  -2 (\mathbf{P}_{\pi}\vec{\mathbf{y}})^T  \mathbf{t}+\| \mathbf{t}\|^2 }{2 \sigma^2}},
\end{align*}
and canceling the same terms. 
\end{IEEEproof} 
In Fig.~\ref{fig:MMSEcurves},
the performance of the MLE (dashed curve) in Lemma~\ref{lem:MLEGaussi} is compared to the other two suboptimal estimators, and the optimal estimator.   
From Fig.~\ref{fig:MMSEcurves} it appears that the MLE is a suitable estimator only when $\sigma \to 0$. The reason is that when $\sigma \to 0$, it can be shown that  $  {\bf  \hat{\boldsymbol{\ell}}}_{\text{MLE}}( \vec{\mathbf{y}}) =\vec{\mathbf{y}}$, which is an optimal estimator of $\vec{\mathbf{X}}$ since $\lim_{\sigma \to 0}   \vec{\mathbf{Y}}=\vec{\mathbf{X}}$ almost surely.

\section{Large-Dimension Regime}
\label{sec:HighN}

In this section, we analyze the behavior of the MMSE of estimating $\vec{\mathbf{X}}$ from an observation $\vec{\mathbf{Y}}$ as $n \to \infty$.  
In particular, our main result is provided in the next theorem, whose proof is presented throughout this section.
\begin{thm} 
\label{thm:HighNVar}
Let $\X \sim \mathcal{N}(\mathbf{0}_n, \mathbf{I}_n)$. Then, we have
\begin{equation*}
\lim_{n \to \infty} \frac{\mathsf{Var}(\vec{\mathbf{X}})}{n}=0.
\end{equation*}
\end{thm}
The behavior of $\mathsf{Var}(\vec{\mathbf{X}})/n$ versus different values of $n$ is shown in  Fig.~\ref{fig:Var}.
From Theorem~\ref{thm:HighNVar}, we can readily obtain the behavior of the MMSE as $n \to \infty$, which is stated next.
\begin{cor}
\label{cor:MMSELargen}
Let $\X \sim \mathcal{N}(\mathbf{0}_n, \mathbf{I}_n)$, and  let $\Y| \X=\mathbf{x} \sim \mathcal{N}({\bf x}, \sigma^2  \mathbf{I}_n)$. Then, we have
\begin{equation*}
\lim_{n \to \infty} \frac{\mmse(\vec{\mathbf{X}}|\vec{\mathbf{Y}})}{n} =\lim_{n \to \infty} \frac{\mmse(\vec{\mathbf{X}}|{\mathbf{Y}})}{n} =0.
\end{equation*}
\end{cor}
\begin{IEEEproof}
First note that
\begin{equation*}
\min \left \{\mmse(\vec{\mathbf{X}}|\vec{\mathbf{Y}}), \mmse(\vec{\mathbf{X}}|{\mathbf{Y}}) \right \} \geq 0.
\end{equation*}
Moreover, from the definition of MMSE, we have that
\begin{equation*}
\mmse(\vec{\mathbf{X}}|\vec{\mathbf{Y}}) 
= \mathbb{E} \left [ \left \| \vec{\mathbf{X}} - \mathbb{E} \left [ \vec{\mathbf{X}} | \vec{\mathbf{Y}}\right ] \right \|^2 \right ]
%
\leq \mathsf{Var}(\vec{\mathbf{X}}),
\end{equation*}
where the inequality follows by using $\mathbb{E} \left [ \vec{\mathbf{X}}\right ]$ as suboptimal estimator. This implies that 
\begin{equation*}
\lim_{n \to \infty} \frac{\mmse(\vec{\mathbf{X}}|\vec{\mathbf{Y}})}{n} =0.
\end{equation*}
By replacing $\vec{\mathbf{Y}}$ with ${\mathbf{Y}}$, 
it also follows that 
\begin{equation*}
\lim_{n \to \infty} \frac{\mmse(\vec{\mathbf{X}}|{\mathbf{Y}})}{n} =0.
\end{equation*}
This concludes the proof of Corollary~\ref{cor:MMSELargen}.
\end{IEEEproof}
\begin{rem}
Corollary~\ref{cor:MMSELargen} implies that the MMSE of estimating $\vec{\mathbf{X}}$ from an observation $\vec{\mathbf{Y}}$ grows sublinearly in $n$. This is significantly different from the MMSE of estimating ${\mathbf{X}}$ from an observation ${\mathbf{Y}}$, which grows linearly with $n$, namely~\cite{kay1993fundamentals}
\begin{equation*}
\mmse({\mathbf{X}}|{\mathbf{Y}}) = n \frac{\sigma^2}{1+\sigma^2}.
\end{equation*}
\end{rem}
To prove Theorem~\ref{thm:HighNVar}, we next derive novel approximations on the first moment and variance of order statistics, which are tight when $n \to \infty$. 
We start by proposing the following lower bound on the variance, which will be used
in Section~\ref{sec:LowerBoundsOnMMSE}.  

\begin{thm}\label{thm:FistBoundOnTHevariance} For any $\vec{\X}$, we have that
\begin{equation}
 \mathsf{Var}(\vec{\mathbf{X}})  \ge  \mathsf{Var}(\| \X\|).  \label{eq:LowerBoundOnVar}
\end{equation} 
In particular, for $\X \sim \mathcal{N}({\bf 0}_n,   \mathbf{I}_n)$ we have the following:
\begin{itemize}
\item  For every $n \ge 1$
\begin{equation}
\label{eq:Prop1Norm}
 \mathsf{Var}(\| \X\|)= n -2 \left( \frac{  \Gamma \left(  \frac{n+1}{2}\right)}{ \Gamma \left(  \frac{n}{2}\right)} \right)^2,
\end{equation}
where $\Gamma (\cdot)$ is the gamma function; and
\item  $  n \to \mathsf{Var}(\| \X\|)$ is monotonically increasing with  
\begin{equation}
\label{eq:Prop2Norm}
\lim_{n \to \infty} \mathsf{Var}(\| \X\|) = \frac{1}{2}.
\end{equation}
\end{itemize}
\end{thm}
\begin{IEEEproof}
The proof of the lower bound in \eqref{eq:LowerBoundOnVar} follows by using modulus inequality
\begin{align*}
 \mathsf{Var}(\vec{\mathbf{X}}) &= \E[ \| \vec{\mathbf{X}}\|^2]- \| \E[  \vec{\mathbf{X}} ] \|^2 \\
 &\ge   \E[ \| \vec{\mathbf{X}}\|^2]-  (\E[  \| \vec{\mathbf{X}} \| ] )^2\\
 &= \mathsf{Var}( \| \vec{\mathbf{X}} \|) = \mathsf{Var}(\|\mathbf{X}\|).  
\end{align*} 
The proofs of~\eqref{eq:Prop1Norm} and~\eqref{eq:Prop2Norm} follow from the fact  that   for $\X \sim \mathcal{N}({\bf 0}_n,   \mathbf{I}_n)$, then $\| \X\|$ is distributed according to a chi distribution for which closed-form expressions and other properties of  the variance can be found in~\cite[Chapter~8]{forbes2011statistical}. 
\end{IEEEproof}
%
Next, we give an approximation for the variance that is accurate within $\sqrt{n \log(n)}$, the proof of which is in Appendix~\ref{app:OrdStatMomVar}.
%
%
\begin{thm}
\label{thm:boundForVar}
Let $\X \sim \mathcal{N}({\bf 0}_n,   \mathbf{I}_n)$. Then,
\begin{align}
& \left | \E[X_{(i)}]  -  \Phi^{-1} \left ( \frac{i}{n+1} \right ) \right |  \nonumber
\\  \le &     \sqrt{ \frac{\pi}{2} } \sqrt{  \frac{2}{i} + \frac{2}{n+1-i}  +  \left |   \frac{1}{i}- \frac{1}{2(n+1-i)}  \right|^2 },
\label{eq:ExpOrdStat}
\end{align}
and
\begin{align}
& \left| \mathsf{Var}(\vec{\mathbf{X}}) - \left(  n - \sum_{i=1}^n  \left( \Phi^{-1} \left ( \frac{i}{n+1} \right )  \right)^2\right) \right| \notag
\\  \le & 2  \sqrt{ 18 (n+1) \frac{3\pi}{2} \left( 2 \log(n) +  \frac{1}{n}  +1\right) } \notag
\\& \quad +  \frac{3\pi}{2} \left( 2 \log(n) +  \frac{1}{n}  +1\right) ,
\label{eq:VarOrdStat}
 \end{align}
where $\Phi^{-1}(\cdot)$ is the inverse cumulative distribution function of the normal distribution (i.e., quantile function).
\end{thm}
\begin{rem}
\label{rem:AltBoundVar}
Note that, by inspection, we have that
\begin{align*}
 \E \left[ \left |  \Phi^{-1}(U_n)   \right|^2    \right] = \frac{1}{n} \sum_{i=1}^n  \left( \Phi^{-1} \left ( \frac{i}{n+1} \right )  \right)^2,
 \end{align*} 
where $U_n$ is an uniform random variable distributed on $\{ \frac{1}{n+1}, \ldots \frac{n}{n+1} \}$.
This alternative expression can be used, for example, in the bound in~\eqref{eq:VarOrdStat}. 
\end{rem}
To conclude the proof of Theorem~\ref{thm:HighNVar}, we require the following technical lemma whose proof can be found in Appendix~\ref{lem:LimExp}.
  \begin{lem} \label{lem:ExchangeOflImit}  
Let $U_n$ be uniformly distributed on $\{ \frac{1}{n+1}, \ldots \frac{n}{n+1} \}$ and let $U$ be uniform on $(0,1)$. Then,  
  \begin{align*}
   \lim_{n \to \infty} \E \left[ \left |  \Phi^{-1}(U_n)   \right|^2    \right]=   \E \left[ \left |  \Phi^{-1}(U)   \right|^2    \right]. 
  \end{align*}
  \end{lem} 
As a consequence of Theorem~\ref{thm:boundForVar}, Remark~\ref{rem:AltBoundVar} and Lemma~\ref{lem:ExchangeOflImit}, we obtain
\begin{align}
 \lim_{n \to \infty} \frac{1}{n } \mathsf{Var}(\vec{\mathbf{X}})  &=  1- \lim_{n \to \infty}  \E \left[ \left |  \Phi^{-1}(U_n)   \right|^2    \right] \notag \\
 &= 1-   \E \left[ \left |  \Phi^{-1}(U)   \right|^2    \right] \notag \\
 &\stackrel{{\rm{(a)}}}{=}1-   \E \left[ X^2    \right]=1-1=0, \notag
\end{align}
where the equality in $\rm{(a)}$ follows since $X \stackrel{d}{=}\Phi^{-1}(U)$.
This concludes the proof of Theorem~\ref{thm:HighNVar}.

\section{Discussion on Lower Bounds}
\label{sec:LowerBoundsOnMMSE}
In this section, we seek to further analyze the expression of $\mmse(\vec{\mathbf{X}}|\vec{\mathbf{Y}})$. Towards this end, we focus on lower bounds on the MMSE, with the goal to asses fundamental limits of recovering  $\vec{\mathbf{X}}$ from $\vec{\mathbf{Y}}$.
%
%
%
In particular, we discuss two commonly used lower bounds, namely the Bayesian Cram\'er-Rao bound and the maximum entropy bound. Surprisingly, and somewhat disappointingly, we show that both these commonly used lower bounds are unsuitable for our purposes.    
This opens up  new research avenues of deriving tight lower bounds on $\mmse(\vec{\mathbf{X}}|\vec{\mathbf{Y}})$, which is object of current investigation.

\subsection{Bayesian Cram\'er-Rao Bound}
We here focus on the Bayesian Cram\'er-Rao bound. An important thing to note about the Bayesian Cram\'er-Rao is that this bound holds under the assumption of the following regularity  condition \cite{weinstein1988general}: 
\begin{align}
\E \left[ \phi(\vec{\mathbf{X}}, \vec{\mathbf{Y}}  )  | \vec{\mathbf{Y}} = \vec{\mathbf{y}}   \right]=\mathbf{0}_n,  \forall  \vec{\mathbf{y}},  \label{eq:RegularityConditionForBCR}
\end{align}
where
\begin{align}
\phi(\vec{\mathbf{x}}, \vec{\mathbf{y}}  )=\nabla_{\vec{\mathbf{x}}}  \log f_{\vec{\mathbf{X}}, \vec{\mathbf{Y}} } (\vec{\mathbf{x}} ,\vec{\mathbf{y}} ). 
\end{align}
We next show that the regularity condition in \eqref{eq:RegularityConditionForBCR} does not hold and, hence, the Bayesian Cram\'er-Rao bound cannot be used to lower bound  $\mmse(\vec{\mathbf{X}}|\vec{\mathbf{Y}})$. 

\begin{lem} 
\label{lem:CRn2}
Let $\X \sim \mathcal{N}(\mathbf{0}_n, \mathbf{I}_n)$ and  let $\Y| \X=\mathbf{x} \sim \mathcal{N}({\bf x}, \sigma^2  \mathbf{I}_n)$ with $\sigma=1$. Then, for $n=2$  there exists an open $S \subset  \vec{\mathbb{R}}_n$ such that 
\begin{align*}
 \E \left[ \phi(\vec{\mathbf{X}}, \vec{\mathbf{Y}}  )  | \vec{\mathbf{Y}} = \vec{\mathbf{y}}   \right]  \neq  \mathbf{0}_n  ,  \vec{\mathbf{y}} \in S. 
\end{align*}
\end{lem} 
\begin{IEEEproof}
We start by noting that
\begin{align}
\phi(\vec{\mathbf{x}}, \vec{\mathbf{y}}  )&=\nabla_{\vec{\mathbf{x}}}  \log f_{\vec{\mathbf{X}}, \vec{\mathbf{Y}} } (\vec{\mathbf{x}} ,\vec{\mathbf{y}} )= \frac{  \nabla_{\vec{\mathbf{x}}}  f_{\vec{\mathbf{X}}, \vec{\mathbf{Y}} } (\vec{\mathbf{x}} ,\vec{\mathbf{y}} )}{ f_{\vec{\mathbf{X}}, \vec{\mathbf{Y}} } (\vec{\mathbf{x}} ,\vec{\mathbf{y}} )} .\label{eq:TakingGradient}
\end{align}
Therefore,
\begin{align*}
&\E \left[ \phi(\vec{\mathbf{X}}, \vec{\mathbf{Y}}  )  | \vec{\mathbf{Y}} = \vec{\mathbf{y}}   \right] \notag\\
 \stackrel{{\rm{(a)}}}{=} &  \int  \frac{  \nabla  f_{\vec{\mathbf{X}}, \vec{\mathbf{Y}} } (\vec{\mathbf{x}} ,\vec{\mathbf{y}} )}{ f_{\vec{\mathbf{X}}, \vec{\mathbf{Y}} } (\vec{\mathbf{x}} ,\vec{\mathbf{y}} )}   f_{\vec{\mathbf{X}}| \vec{\mathbf{Y}} } (\vec{\mathbf{x}} |\vec{\mathbf{y}} )   {\rm d}  \vec{\mathbf{x}} \notag \\
\stackrel{{\rm{(b)}}}{=} & \frac{1}{  f_{ \vec{\mathbf{Y}} } (\vec{\mathbf{y}} )} \int  \nabla  f_{\vec{\mathbf{X}}, \vec{\mathbf{Y}} } (\vec{\mathbf{x}} ,\vec{\mathbf{y}} )    {\rm d}  \vec{\mathbf{x}} \notag \\
\stackrel{{\rm{(c)}}}{=} & \frac{n!}{  f_{ \vec{\mathbf{Y}} } (\vec{\mathbf{y}} )}  \sum_{ \pi \in  \mathcal{P} }  \int_{\vec{\mathbb{R}}_n}  \nabla \left( f_{\mathbf{Y}|\mathbf{X}} (\mathbf{P}_{\pi}\vec{\mathbf{y}}  |\mathbf{x} )  f_{\mathbf{X} } (\mathbf{x}) \right)   {\rm d}  \mathbf{x} \notag \\
%
\stackrel{{\rm{(d)}}}{=} & \frac{n!}{  f_{ \vec{\mathbf{Y}} } (\vec{\mathbf{y}} )} \sum_{ \pi \in  \mathcal{P} }  \int_{\vec{\mathbb{R}}_n}  (\mathbf{P}_{\pi}\vec{\mathbf{y}}  -2\mathbf{x} )  f_{\mathbf{Y}|\mathbf{X}} (\mathbf{P}_{\pi}\vec{\mathbf{y}}  |\mathbf{x} )  f_{\mathbf{X} } ({\mathbf{x}})   {\rm d}  \mathbf{x}, 
\end{align*} 
where the labeled equalities follow from:   
${\rm{(a)}}$ the definition of expectation and using~\eqref{eq:TakingGradient};  
${\rm{(b)}}$ using Bayes' rule;  
${\rm{(c)}}$ using Lemma~\ref{lem:Lemma_exchangeable} and Lemma~\ref{lem:forMAP}; and  ${\rm{(d)}}$ taking the gradient.

Now, since the above expression is an integral of  Gaussian functions on $\vec{\mathbb{R}}_n$, it follows that   $ \vec{\mathbf{y}}  \to \E \left[ \phi(\vec{\mathbf{X}}, \vec{\mathbf{Y}}  )  | \vec{\mathbf{Y}} = \vec{\mathbf{y}}   \right] $ is a continuous function on $\vec{\mathbb{R}}_n$.  Thus, by continuity, if we can demonstrate that
\begin{align}
 \E \left[ \phi(\vec{\mathbf{X}}, \vec{\mathbf{Y}}  )  | \vec{\mathbf{Y}} =  \mathbf{0}_n   \right]  \neq  \mathbf{0}_n, \label{eq:TestTODO}
\end{align} 
then 
\begin{align}
 \E \left[ \phi(\vec{\mathbf{X}}, \vec{\mathbf{Y}}  )  | \vec{\mathbf{Y}} = \vec{\mathbf{y}}   \right]  \neq  \mathbf{0}_n,
\end{align} 
in some open subset of $\vec{\mathbb{R}}_n$. This would violate the condition in~\eqref{eq:RegularityConditionForBCR}. 
By setting $\vec{\mathbf{y}}= \mathbf{0}_n$, we have that
\begin{align}
&\E \left[ \phi(\vec{\mathbf{X}}, \vec{\mathbf{Y}}  )  | \vec{\mathbf{Y}} =   \mathbf{0}_n  \right]\notag\\
=& -2\frac{(n!)^2}{  f_{ \vec{\mathbf{Y}} } (\mathbf{0}_n )} \int_{\vec{\mathbb{R}}_n}  \mathbf{x}   f_{\mathbf{Y}|\mathbf{X}} (\mathbf{0}_n  |\mathbf{x} )  f_{\mathbf{X} } ({\mathbf{x}})   {\rm d}  \mathbf{x} \notag \\
=& -2n!  ( 4 \pi    )^{ \frac{n}{2} }   \int_{\vec{\mathbb{R}}_n}  \mathbf{x}   f_{\mathbf{Y}|\mathbf{X}} ( \mathbf{0}_n  |\mathbf{x} )  f_{\mathbf{X} } ({\mathbf{x}})   {\rm d}  \mathbf{x}, \label{eq:IntegralOfTestFunction}
\end{align} 
where in the last step we have used that $f_{ \vec{\mathbf{Y}} } (\mathbf{0}_n)= \frac{n!}{( 4 \pi    )^{ \frac{n}{2} }}$ (see Lemma~\ref{lem:Lemma_exchangeable}).

We now consider each of the $n=2$ coordinates of~\eqref{eq:IntegralOfTestFunction}.  
Towards this end, we need to compute the following integral: 
\begin{align*}
\int_{-\infty}^\infty \int_{x_1}^\infty x_i   \frac{1}{ (2 \pi)^2 }  \eu^{- \frac{x_1^2+x_2^2}{2}}   \eu^{- \frac{x_1^2+x_2^2}{2}}    {\rm d}  x_2   {\rm d}  x_1
=(-1)^i\frac{1}{(2 \pi)^{ \frac{3}{2}} 4}, 
\end{align*} 
for $i \in [2]$.
Combining~\eqref{eq:IntegralOfTestFunction} with the integral above, we obtain
\begin{align}
\E \left[ \phi(\vec{\mathbf{X}}, \vec{\mathbf{Y}}  )  | \vec{\mathbf{Y}} = \mathbf{0}_n    \right]= - \frac{2}{\sqrt{2 \pi}} \left [ \begin{array}{l}  -1 \\
1 \end{array} \right].  \label{eq:ValueOFINtevalr}
\end{align} 
The expression in \eqref{eq:ValueOFINtevalr} shows that \eqref{eq:TestTODO} holds. This concludes the proof of Lemma~\ref{lem:CRn2}.  
\end{IEEEproof} 

\subsection{Maximum Entropy Bound}
We here consider the maximum entropy bound on the MMSE~\cite[Chapter~2.2]{elgamalkimbook}, which uses information theoretic arguments and states that for any pair of random vectors $(\mathbf{V},\mathbf{U})$ of dimension $n$, we have that (by also using the upper bound on the determinant of a matrix in terms of the trace)
\begin{align}
\mmse(\mathbf{V}|\mathbf{U}) &\ge    \frac{n}{2 \pi \eu}  \eu^{  \frac{2}{n} h( \mathbf{V}|\mathbf{U} )}   \notag
\\& =  \frac{n}{2 \pi \eu}  \eu^{  \frac{2}{n} h(  \mathbf{V})}    \eu^{ -  \frac{2}{n} I( \mathbf{V};\mathbf{U}  )},   \label{eq:maximumEntropyPrincipleBound}
\end{align}
where $h(  \mathbf{V}),   h( \mathbf{V}|\mathbf{U} )$ and $I( \mathbf{V};\mathbf{U}  )$ are the differential entropy, the conditional differential entropy and the mutual information, respectively.   
The inequality in~\eqref{eq:maximumEntropyPrincipleBound} holds without any assumption on regularity conditions.  
We now use the inequality in \eqref{eq:maximumEntropyPrincipleBound} to derive the following lower bound.

\begin{thm} 
\label{thm:MEB}
Let $\X \sim \mathcal{N}(\mathbf{0}_n, \mathbf{I}_n)$ and  let $\Y| \X=\mathbf{x} \sim \mathcal{N}({\bf x}, \sigma^2  \mathbf{I}_n)$. Then, 
\begin{align}
\mmse(\vec{\mathbf{X}}|\vec{\mathbf{Y}})   \ge \frac{1}{  (n!)^{  \frac{2}{n} } }     \mmse(\mathbf{X}|\mathbf{Y}),      \label{eq:ENTropyBOUNDGaussian}
\end{align} 
where
\begin{align*}
 \mmse(\mathbf{X}|\mathbf{Y})    = n \frac{ \sigma^2}{1+\sigma^2}. 
\end{align*}
Moreover,  as $\sigma \to \infty$ the bound in \eqref{eq:ENTropyBOUNDGaussian} reduces to
\begin{align}
\mathsf{Var}(\vec{\mathbf{X}})  \ge \frac{n}{2 \pi \eu}  \eu^{  \frac{2}{n} h(  \vec{\mathbf{X}})} =   \frac{n}{  (n!)^{  \frac{2}{n} } } \approx \frac{1}{n^{\frac{n+1}{n}}}. \label{eq:varianceLowerBound}
\end{align} 
\end{thm} 
\begin{IEEEproof}
Specializing the bound in \eqref{eq:maximumEntropyPrincipleBound} to our setting we have the following inequality:
\begin{align}
\mmse(\vec{\mathbf{X}}|\vec{\mathbf{Y}})  &\ge   \frac{n}{2 \pi \eu}  \eu^{  \frac{2}{n} h(  \vec{\mathbf{X}})}    \eu^{ -  \frac{2}{n} I(  \vec{\mathbf{X}}; \vec{\mathbf{Y}}   )} . \label{eq:EPIforOurexample}
\end{align} 
We first compute $h(  \vec{\mathbf{X}}) $  in \eqref{eq:EPIforOurexample}, and we obtain
\begin{align}
h(  \vec{\mathbf{X}})& = - \int_{\vec{\mathbb{R}}_n} f_{ \vec{\mathbf{X}}} (  \mathbf{x} ) \log (  f_{ \vec{\mathbf{X}}} (  \mathbf{x} )   ) {\rm d} \mathbf{x}  \notag \\
& \stackrel{{\rm{(a)}}}{=}  - \int_{\vec{\mathbb{R}}_n} f_{ \vec{\mathbf{X}}} (  \mathbf{x} ) \log (   n! f_{ {\mathbf{X}}} (  \mathbf{x} )   )  {\rm d} \mathbf{x} \notag \\
&\stackrel{{\rm{(b)}}}{=}  - \int_{\vec{\mathbb{R}}_n} f_{ \vec{\mathbf{X}}} (  \mathbf{x} ) \log \left(      \frac{n!}{ ( 2 \pi)^{  \frac{n}{2}  }}  \eu^{-  \frac{\| \mathbf{x}  \|^2}{2}} \right ) {\rm d} \mathbf{x} \notag \\
&= - \log \left(      \frac{n!}{ ( 2 \pi)^{  \frac{n}{2}  }}   \right ) +  \int_{\vec{\mathbb{R}}_n} f_{ \vec{\mathbf{X}}} (  \mathbf{x} )  \frac{\| \mathbf{x}  \|^2}{2} {\rm d} \mathbf{x} \notag \\
&\stackrel{{\rm{(c)}}}{=} - \log \left(      \frac{n!}{ ( 2 \pi)^{  \frac{n}{2}  }}   \right ) +    \frac{n }{2} \notag \\
&=   \frac{n }{2} \log \left(    2 \pi \eu   \right )   - \log(n!), \label{eq:EntropyOfSortedX}
\end{align} 
where the labeled equalities follow from:  
${\rm{(a)}}$ Lemma~\ref{lem:Lemma_exchangeable};  ${\rm{(b)}}$ using that $ f_{{\mathbf{X}}}(\mathbf{x})=\frac{1}{ ( 2 \pi)^{  \frac{n}{2}  }}  \eu^{-  \frac{\| \mathbf{x}  \|^2}{2}}$; 
and ${\rm{(c)}}$ using the fact that $\int_{\vec{\mathbb{R}}_n} f_{ \vec{\mathbf{X}}} (  \mathbf{x} )  \| \mathbf{x}  \|^2 {\rm d} \mathbf{x}= \E \left[ \| \vec{\mathbf{X}}\|^2 \right]=n$. 

Next, we further upper bound   $ I(  \vec{\mathbf{X}}; \vec{\mathbf{Y}}   )$ in \eqref{eq:EPIforOurexample}. This is done by using the  data processing inequality  in view of the Markov chain $  \vec{\mathbf{X}} \to \X \to \Y \to \vec{\mathbf{Y}}$, that  is
\begin{align}
 I(  \vec{\mathbf{X}}; \vec{\mathbf{Y}}   ) \le    I(  \mathbf{X};  \mathbf{Y}   )=   \frac{n}{2} \log \left(1 + \frac{1}{ \sigma^2} \right),  \label{eq:ApplyinDataProcessing}
\end{align}  
where the equality follows from the closed-form expression for the mutual information of Gaussian random vectors~\cite[Chapter~9]{Cover:InfoTheory}.
Thus, combing~\eqref{eq:EPIforOurexample},~\eqref{eq:EntropyOfSortedX} and~\eqref{eq:ApplyinDataProcessing} concludes the proof of~\eqref{eq:ENTropyBOUNDGaussian} in Theorem~\ref{thm:MEB}.

We now prove~\eqref{eq:varianceLowerBound} in Theorem~\ref{thm:MEB}. To this end, we take  $\sigma  \to \infty$ on both side of~\eqref{eq:ENTropyBOUNDGaussian}, which leads to 
\begin{equation*}
\mathsf{Var}(\vec{\mathbf{X}})  \ge   \frac{n}{  (n!)^{  \frac{2}{n} } } \approx  \frac{1}{n^{\frac{n+1}{n}}},
\end{equation*}
where we have used that  $ \lim_{\sigma  \to \infty} \mmse(\vec{\mathbf{X}}|\vec{\mathbf{Y}}) =\mathsf{Var}(\vec{\mathbf{X}}) $ (see Theorem~\ref{thm:EstPerHighNoise}) and the approximation is due to Stirling's formula.
This concludes the proof of Theorem~\ref{thm:MEB}.
\end{IEEEproof}

The expression in \eqref{eq:ENTropyBOUNDGaussian} is pleasing in its simplicity and, hence, it would be interesting to evaluate the effectiveness of this bound.    In particular, it would be helpful to understand how the bound in \eqref{eq:ENTropyBOUNDGaussian} scales as $n$ tends to infinity.    
However, after closer examination, for $n \to \infty$ by comparing the lower bound in~\eqref{eq:varianceLowerBound} with the lower bound on $\mathsf{Var}(\vec{\mathbf{X}})$  in~\eqref{eq:LowerBoundOnVar}, it is apparent that the bound in~\eqref{eq:varianceLowerBound}, which decreases with $n$, does not scale properly as $n$ tends to infinity.  


\begin{appendices}

\section{Auxiliary Results on the Distribution of $(\vec{\mathbf{X}}, \vec{\mathbf{Y}})$}
\label{app:AuxRes}
 In this section, we state and prove some properties on exchangeable random variables.
\begin{lem}
\label{lem:Lemma_exchangeable}
Let $\mathbf{X}$ be an exchangeable random vector. Then,
\begin{align}
f_{\vec{\mathbf{X}}|\mathbf{X}} (\vec{\mathbf{x}}| \mathbf{x}) =  \sum_{ \pi \in  \mathcal{P} }  \delta({\mathbf{x}}-\mathbf{P}_{\pi}\vec{\mathbf{x}}), \label{eq:deltaX}
\end{align}
and 
\begin{equation}
f_{\vec{\mathbf{X}}}(\vec{\mathbf{x}}) =n!  f_{\mathbf{X}}(\vec{\mathbf{x}}).
\label{eq:exch_dist}
\end{equation}
\end{lem}

\begin{IEEEproof} The proof of \eqref{eq:deltaX} follows by inspection. 

To show \eqref{eq:exch_dist} observe that 
\begin{align*}
f_{\vec{\mathbf{X}}} (\vec{\mathbf{x}}) & = \int_{\vec{\mathbb{R}}_n}  f_{\vec{\mathbf{X}}|\mathbf{X}} (\vec{\mathbf{x}}| \mathbf{x}) f_{\mathbf{X}}(\mathbf{x})  \ {\rm d} \mathbf{x}
\\ & \stackrel{{\rm{(a)}}}{=}  \int_{\vec{\mathbb{R}}_n}  f_{\mathbf{X}}(\mathbf{x}) \sum_{ \pi \in  \mathcal{P} }
\delta({\mathbf{x}}-\mathbf{P}_{\pi}\vec{\mathbf{x}}) \ {\rm d} \mathbf{x}
\\& \stackrel{{\rm{(b)}}}{=} \sum_{ \pi \in  \mathcal{P} } f_{\mathbf{X}} (\mathbf{P}_{\pi}\vec{\mathbf{x}})
\\ & \stackrel{{\rm{(c)}}}{=} n! f_{\mathbf{X}}(\vec{\mathbf{x}}),
\end{align*}
where the labeled equalities follow from:  $\rm{(a)}$ using~\eqref{eq:deltaX};
$\rm{(b)}$ the sifting property of the delta function; and $\rm{(c)}$ the exchangeable property of $\mathbf{X}$.
This concludes the proof of Lemma~\ref{lem:Lemma_exchangeable}.
\end{IEEEproof}

\begin{lem}
\label{lem:forMAP}
Let  $\mathbf{X}$ be an exchangeable random vector (see Definition~\ref{def:ExchRanVar}). Then, 
\begin{align}
f_{\vec{\mathbf{Y}} |  \vec{\mathbf{X}} }( \vec{\mathbf{y}} | \vec{\mathbf{x}})=\sum_{ \pi \in  \mathcal{P} }  f_{\mathbf{Y}|\mathbf{X} } (\mathbf{P}_{\pi}\vec{\mathbf{y}} | \vec{\mathbf{x}}), 
\label{eq:sortingChannelProof}   
\end{align} 
and 
\begin{align}
f_{\vec{\mathbf{X}}|\vec{\mathbf{Y}} }( \vec{\mathbf{x}} | \vec{\mathbf{y}}) = \frac{n!}{f_{\vec{\mathbf{Y}} }(\vec{\mathbf{y}})}  f_{\mathbf{X}}(\vec{\mathbf{x}})
\sum_{ \pi \in  \mathcal{P} }  f_{\mathbf{Y}|\mathbf{X} } (\mathbf{P}_{\pi}\vec{\mathbf{y}} | \vec{\mathbf{x}}).  \label{eq:conditonalForAposteriory}
\end{align}
\end{lem}
\begin{IEEEproof}
To show~\eqref{eq:sortingChannelProof}, observe that
\begin{align*}
&f_{\vec{\mathbf{Y}} |  \vec{\mathbf{X}} }( \vec{\mathbf{y}} | \vec{\mathbf{x}}) \notag\\
  \stackrel{{\rm{(a)}}}{=} & \int_{\vec{\mathbb{R}}_n} f_{\vec{\mathbf{Y}} |  \mathbf{X} }( \vec{\mathbf{y}} | \mathbf{x})  f_{\mathbf{X}|\vec{\mathbf{X}}}(  \mathbf{x}|\vec{\mathbf{x}} ) \ {\rm d} \mathbf{x} \\
%
%
\stackrel{{\rm{(b)}}}{=} & \frac{1}{f_{\vec{\mathbf{X}} }(\vec{\mathbf{x}})}
 \int_{\vec{\mathbb{R}}_n} f_{\vec{\mathbf{Y}} |  \mathbf{X} }( \vec{\mathbf{y}} | \mathbf{x}) f_{\vec{\mathbf{X}}|\mathbf{X}} ( \vec{\mathbf{x}} |\mathbf{x}) f_{\mathbf{X}}(\mathbf{x}) \ {\rm d} \mathbf{x}  \\
  \stackrel{{\rm{(c)}}}{=} &  \frac{1}{f_{\vec{\mathbf{X}} }(\vec{\mathbf{x}})} \int_{\vec{\mathbb{R}}_n}  f_{\vec{\mathbf{X}}|\mathbf{X}} ( \vec{\mathbf{x}}  |\mathbf{x}) f_{\mathbf{X}}(\mathbf{x}) 
\\& \qquad \qquad \left(\int_{\vec{\mathbb{R}}_n}  f_{\vec{\mathbf{Y}} |  \mathbf{Y} }( \vec{\mathbf{y}} | \mathbf{y}) f_{\mathbf{Y}|\mathbf{X} } (\mathbf{y} | \mathbf{x}) \ {\rm d} \mathbf{y}  \right) {\rm d} \mathbf{x}   \\
 \stackrel{{\rm{(d)}}}{=} &  \frac{1}{f_{\vec{\mathbf{X}} }(\vec{\mathbf{x}})} \int_{\vec{\mathbb{R}}_n}  f_{\vec{\mathbf{X}}|\mathbf{X}} (\vec{\mathbf{x}}|\mathbf{x}) f_{\mathbf{X}}(\mathbf{x}) \sum_{ \pi \in  \mathcal{P} }  f_{\mathbf{Y}|\mathbf{X} } (\mathbf{P}_{\pi}\vec{\mathbf{y}} | \mathbf{x}) \   {\rm d} \mathbf{x}   \\
  \stackrel{{\rm{(e)}}}{=} & 
\frac{1}{f_{\vec{\mathbf{X}} }(\vec{\mathbf{x}})}   \sum_{ \pi' \in  \mathcal{P} }f_{\mathbf{X}}( \mathbf{P}_{\pi'} \vec{\mathbf{x}}) \sum_{ \pi \in  \mathcal{P} }  f_{\mathbf{Y}|\mathbf{X} } (\mathbf{P}_{\pi}\vec{\mathbf{y}} | \mathbf{P}_{\pi'} \vec{\mathbf{x}})   \\
 \stackrel{{\rm{(f)}}}{=} & \frac{1}{f_{\vec{\mathbf{X}} }(\vec{\mathbf{x}})}    f_{\mathbf{X}}(\vec{\mathbf{x}}) \sum_{ \pi' \in  \mathcal{P} }\sum_{ \pi \in  \mathcal{P} }  f_{\mathbf{Y}|\mathbf{X} } (\mathbf{P}_{\pi}\vec{\mathbf{y}} | \mathbf{P}_{\pi'} \vec{\mathbf{x}})   \\
 \stackrel{{\rm{(g)}}}{=}  & \frac{1}{f_{\vec{\mathbf{X}} }(\vec{\mathbf{x}})}  n! f_{\mathbf{X}}(\vec{\mathbf{x}})
\sum_{ \pi \in  \mathcal{P} }  f_{\mathbf{Y}|\mathbf{X} } (\mathbf{P}_{\pi}\vec{\mathbf{y}} | \vec{\mathbf{x}})    \\
 \stackrel{{\rm{(h)}}}{=} & \sum_{ \pi \in  \mathcal{P} }  f_{\mathbf{Y}|\mathbf{X} } (\mathbf{P}_{\pi}\vec{\mathbf{y}} | \vec{\mathbf{x}})    ,
\end{align*}
where the labeled equalities follow from: $\rm{(a)}$ using the law of total probability and the Markov chain $ \vec{\mathbf{X}}  \to \X \to \vec{\mathbf{Y}} $; $\rm{(b)}$ using Bayes' rule; $\rm{(c)}$ using the  law of total probability and the Markov chain $  \X \to \Y \to  \vec{\mathbf{Y}}$; $\rm{(d)}$ using the fact that  (see~\eqref{eq:deltaX} in Lemma~\ref{lem:Lemma_exchangeable})
\begin{align}
f_{\vec{\mathbf{Y}}|\mathbf{Y}} (\vec{\mathbf{y}}| \mathbf{y}) = \sum_{ \pi \in  \mathcal{P} } \delta({\mathbf{y}}-\mathbf{P}_{\pi}\vec{\mathbf{y}}), \label{eq:deltaY}
\end{align}
and the sifting property of the delta function;
$\rm{(e)}$ using~\eqref{eq:deltaX} in Lemma~\ref{lem:Lemma_exchangeable} and the sifting property of the delta function;
$\rm{(f)}$ using the assumption of $\X$ being an exchangeable random vector; $\rm{(g)}$    using the assumption in \eqref{eq:PermuationSymmetryCondition}; and
$\rm{(h)}$  using~\eqref{eq:exch_dist} in Lemma~\ref{lem:Lemma_exchangeable}.
The proof of \eqref{eq:conditonalForAposteriory} now follows by using Bayes' rule, the exchangeable property of $\X$  and~\eqref{eq:sortingChannelProof}; we obtain
\begin{align*}
   f_{\vec{\mathbf{X}}|\vec{\mathbf{Y}} }( \vec{\mathbf{x}} | \vec{\mathbf{y}}) 
= \frac{n!}{f_{\vec{\mathbf{Y}} }(\vec{\mathbf{y}})}   f_{\mathbf{X}}(\vec{\mathbf{x}})
\sum_{ \pi \in  \mathcal{P} }  f_{\mathbf{Y}|\mathbf{X} } (\mathbf{P}_{\pi}\vec{\mathbf{y}} | \vec{\mathbf{x}})    .
   \end{align*}
 This concludes the proof of Lemma~\ref{lem:forMAP}.
\end{IEEEproof}

%
%

\section{Proof of Theorem~\ref{thm:ErrorForApproximateEstiamtor}} 
\label{app:proof:thm:ErrorForApproximateEstiamtor}

Using~\eqref{eq:EstimatorBothXandYsymmetric},~\eqref{eq:ProposeSubOptimalEstimator} and modulus inequality, we have that 
\begin{align*}
&
\left|  \mathbb{E}[X_{(k)}| \vec{\mathbf{Y}}=\vec{\mathbf{y}}]-      \hat{f}_k( \vec{\mathbf{y}})  \right |   \notag\\
 \le & 
\sum_{ \pi \in  \mathcal{P} }  \Big |  \mathbb{E} \left[ X_k  \cdot 1_{\vec{\mathbb{R}}_n}(\mathbf{X}) |  \mathbf{Y}=\mathbf{P}_{\pi}\vec{\mathbf{y}}    \right]  \notag\\
& \qquad - \mathbb{E} \left[ X_k   |       \mathbf{Y}=\mathbf{P}_{\pi}\vec{\mathbf{y}}    \right]  \cdot  p_{\vec{\mathbf{X}}|  \mathbf{Y}}( \mathbf{P}_{\pi}\vec{\mathbf{y}}  ) \Big |,  
\end{align*}
 where we let
\begin{align*}
p_{\vec{\mathbf{X}}|  \mathbf{Y}}( \mathbf{P}_{\pi}\vec{\mathbf{y}}) = \mathbb{P} \left[  \mathbf{X} \in \vec{\mathbb{R}}_n |  \mathbf{Y}=\mathbf{P}_{\pi}\vec{\mathbf{y}}      \right].
\end{align*}
Next, observe that
\begin{align}
 & \Big|   \mathbb{E} \left[ X_k \cdot  1_{\vec{\mathbb{R}}_n}(\mathbf{X})   |       \mathbf{Y}=\mathbf{P}_{\pi}\vec{\mathbf{y}}    \right]  \notag\\
 &- \mathbb{E} \left[ X_k   |       \mathbf{Y}=\mathbf{P}_{\pi}\vec{\mathbf{y}}    \right]  \cdot  p_{\vec{\mathbf{X}}|  \mathbf{Y}}( \mathbf{P}_{\pi}\vec{\mathbf{y}}  )  \Big | \notag \\
  \stackrel{{\rm{(a)}}}{=} & \Big|  \mathbb{E} \left[ X_k\cdot 1_{\vec{\mathbb{R}}_n}(\mathbf{X})   |       \mathbf{Y}=\mathbf{P}_{\pi}\vec{\mathbf{y}}    \right] \notag\\
  & - \mathbb{E} \left[ X_k \cdot p_{\vec{\mathbf{X}}|  \mathbf{Y}}( \mathbf{P}_{\pi}\vec{\mathbf{y}}  )   |       \mathbf{Y}=\mathbf{P}_{\pi}\vec{\mathbf{y}}    \right]     \Big| \notag \\
    = & \left|  \mathbb{E} \left[ X_k \cdot \left( 1_{\vec{\mathbb{R}}_n}(\mathbf{X})- p_{\vec{\mathbf{X}}|  \mathbf{Y}}( \mathbf{P}_{\pi}\vec{\mathbf{y}}  )  \right)  \left | \right .       \mathbf{Y}=\mathbf{P}_{\pi}\vec{\mathbf{y}}    \right]    \right| \notag \\
    \stackrel{{\rm{(b)}}}{\le} &  \sqrt{   \mathbb{E} \left[ X_k^2   |       \mathbf{Y}=\mathbf{P}_{\pi}\vec{\mathbf{y}}    \right]  }  \notag\\
    & \quad  \cdot  \sqrt{  \mathbb{E} \left[  \left( 1_{\vec{\mathbb{R}}_n}(\mathbf{X})- p_{\vec{\mathbf{X}}|  \mathbf{Y}}( \mathbf{P}_{\pi}\vec{\mathbf{y}}  )  \right)^2  \left | \right.       \mathbf{Y}=\mathbf{P}_{\pi}\vec{\mathbf{y}}    \right]    } \notag \\
       \stackrel{{\rm{(c)}}}{=} &  \sqrt{   \mathbb{E} \left[ X_k^2   |       \mathbf{Y}=\mathbf{P}_{\pi}\vec{\mathbf{y}}    \right] g (  \mathbf{P}_{\pi}\vec{\mathbf{y}}   )    }  , \label{eq:bound_on_diff_est}
\end{align}
where the labeled (in)-equalities follow from:  $\rm{(a)}$ using the property that $\E[ U f(W) |W]=  \E[ U  |W] f(W)$;   $\rm{(b)}$ using Cauchy-Schwarz inequality;  and  ${\rm{(c)}}$ using 
\begin{align}
&\mathbb{E} \left[  \left( 1_{\vec{\mathbb{R}}_n}(\mathbf{X})- p_{\vec{\mathbf{X}}|  \mathbf{Y}}( \mathbf{P}_{\pi}\vec{\mathbf{y}}  )  \right)^2  |       \mathbf{Y}=\mathbf{P}_{\pi}\vec{\mathbf{y}}    \right]  \notag\\
=&
  \mathbb{E} \left[   1_{\vec{\mathbb{R}}_n}(\mathbf{X})-2 \cdot 1_{\vec{\mathbb{R}}_n}(\mathbf{X}) \cdot  p_{\vec{\mathbf{X}}|  \mathbf{Y}}( \mathbf{P}_{\pi}\vec{\mathbf{y}}  )    |       \mathbf{Y}=\mathbf{P}_{\pi}\vec{\mathbf{y}}    \right]   \notag\\
& \quad +   \mathbb{E} \left[   p_{\vec{\mathbf{X}}|  \mathbf{Y}}^2( \mathbf{P}_{\pi}\vec{\mathbf{y}}  )   |       \mathbf{Y}=\mathbf{P}_{\pi}\vec{\mathbf{y}}    \right] \notag  \\
= & p_{\vec{\mathbf{X}}|  \mathbf{Y}}( \mathbf{P}_{\pi}\vec{\mathbf{y}}  )   \left(   1   - p_{\vec{\mathbf{X}}|  \mathbf{Y}}( \mathbf{P}_{\pi}\vec{\mathbf{y}}  )   \right) 
= g (  \mathbf{P}_{\pi}\vec{\mathbf{y}}   ) \notag,
    \end{align}
 where we used the fact that, for an event $E$, we have $\mathbb{E} \left [ 1_E\right ] = p(E)$ (property of indicator function).

Now, by using the Pythagorean theorem observe that
\begin{align}
&\mathbb{E} \left [ \left \| \vec{\mathbf{X}} -  {\bf  \hat{f}}( \vec{\mathbf{Y}}) \right \|^2 \right ]-  \mmse(\vec{\mathbf{X}}|  \vec{\mathbf{Y}})  \notag\\
=&  \mathbb{E} \left [ \left \|  \mathbb{E}[\vec{\mathbf{X}}| \vec{\mathbf{Y}}]- {\bf  \hat{f}}( \vec{\mathbf{Y}}) \right \|^2 \right ]  \notag\\
=&\sum_{k=1}^n \mathbb{E} \left [ \left |  \mathbb{E}[X_{(k)}| \vec{\mathbf{Y}}]- \hat{f}_k( \vec{\mathbf{Y}})\right |^2 \right ] . \label{eq:PythegorianTheorem}
\end{align} 

Next, using the bound in~\eqref{eq:bound_on_diff_est}  on each term in~\eqref{eq:PythegorianTheorem} we have the following bound:
\begin{align}
&\mathbb{E} \left [ \left |  \mathbb{E}[X_{(k)}| \vec{\mathbf{Y}}]- \hat{f}_k( \vec{\mathbf{Y}})\right |^2 \right ]  \notag\\
 \le &  \sum_{ \pi \in  \mathcal{P} }  \mathbb{E} \left [  \mathbb{E} \left[ X_k^2   |       \mathbf{Y}=\mathbf{P}_{\pi}\vec{\mathbf{Y}}    \right] g( \mathbf{P}_{\pi}\vec{\mathbf{Y}}  )     \right] \notag\\
= & \sum_{ \pi \in  \mathcal{P} } \mathbb{E} \left [  \mathbb{E} \left[ X_k^2   |       \mathbf{Y}=\mathbf{P}_{\pi}\mathbf{Y}      \right] g( \mathbf{P}_{\pi}\mathbf{Y}  )   | \mathbf{Y} \in \vec{\mathbb{R}}_n   \right] \notag\\
\stackrel{{\rm{(a)}}}{=} & \sum_{ \pi \in  \mathcal{P} } \mathbb{E} \left [  \mathbb{E} \left[ X_k^2 \cdot g( \mathbf{P}_{\pi}\mathbf{Y}  )    |       \mathbf{Y}=\mathbf{P}_{\pi}\mathbf{Y}      \right]   | \mathbf{Y} \in \vec{\mathbb{R}}_n   \right]  \notag \\ 
= &  \sum_{ \pi \in  \mathcal{P} } \mathbb{E} \left [  X_k^2 \cdot g \left( \mathbf{P}_{\pi}\mathbf{Y}  \right)      | \mathbf{Y} \in \vec{\mathbb{R}}_n      \right ],\label{eq:BoundOnErrorPerDimension}
\end{align}
where the equality in $\rm{(a)}$ follows by using the property that $\E[ U f(W) |W]=  \E[ U  |W] f(W)$.
Combining \eqref{eq:PythegorianTheorem} and \eqref{eq:BoundOnErrorPerDimension} concludes the proof of Theorem~\ref{thm:ErrorForApproximateEstiamtor}.

\begin{figure*}
\begin{align}
&\mathbb{E} \left [ \left \| \vec{\mathbf{X}} -   {\bf  \hat{h}}( \vec{\mathbf{Y}}_t) \right \|^2 \right ]-  \mmse(\vec{\mathbf{X}}|  \vec{\mathbf{Y}}_t)  \notag\\
%
%
=&\sum_{k=1}^n \mathbb{E} \left [ \left |  \mathbb{E}[X_{(k)}| \vec{\mathbf{Y}}_t]- \hat{h}_k( \vec{\mathbf{Y}}_t)\right |^2 \right ] \notag \\
\stackrel{{\rm{(a)}}}{=}& \sum_{k=1}^n \mathbb{E} \left [ \left |\sum_{ \pi \in  \mathcal{P} } \mathbb{E} \left[ X_k     | \mathbf{X} \in \vec{\mathbb{R}}_n,      \mathbf{Y}_t=\mathbf{P}_{\pi} \vec{\mathbf{Y}}_{t}    \right]     \mathbb{P} \left[  \mathbf{X} \in \vec{\mathbb{R}}_n |  \mathbf{Y}_t=\mathbf{P}_{\pi} \vec{\mathbf{Y}}_{t}      \right] -  \E[ X_k | \X \in  \vec{\mathbb{R}}_n  ] \right |^2 \right ] \notag \\
\stackrel{{\rm{(b)}}}{= }& \sum_{k=1}^n \mathbb{E} \left [ \left |\sum_{ \pi \in  \mathcal{P} } \mathbb{P} \left[  \mathbf{X} \in \vec{\mathbb{R}}_n |  \mathbf{Y}_t=\mathbf{P}_{\pi} \vec{\mathbf{Y}}_{t}      \right] \left ( \mathbb{E} \left[ X_k     | \mathbf{X} \in \vec{\mathbb{R}}_n,      \mathbf{Y}_t=\mathbf{P}_{\pi} \vec{\mathbf{Y}}_{t}    \right]      -  \E[ X_k | \X \in  \vec{\mathbb{R}}_n  ] \right ) \right |^2  \right ] \notag \\
\stackrel{{\rm{(c)}}}{\leq}& n! \sum_{k=1}^n \sum_{ \pi \in  \mathcal{P} } \mathbb{E} \left [ \left | \mathbb{P} \left[  \mathbf{X} \in \vec{\mathbb{R}}_n |  \mathbf{Y}_t=\mathbf{P}_{\pi} \vec{\mathbf{Y}}_{t}      \right] \left ( \mathbb{E} \left[ X_k     | \mathbf{X} \in \vec{\mathbb{R}}_n,      \mathbf{Y}_t=\mathbf{P}_{\pi} \vec{\mathbf{Y}}_{t}    \right]      -  \E[ X_k | \X \in  \vec{\mathbb{R}}_n  ] \right ) \right |^2  \right ] \notag \\
%
%
\stackrel{{\rm{(d)}}}{\leq} & n!  \sum_{k=1}^n  \sum_{ \pi \in  \mathcal{P} } \mathbb{E} \left [ \left |  \E[ X_k | \X \in  \vec{\mathbb{R}}_n, \mathbf{Y}_t=\mathbf{P}_{\pi} \vec{\mathbf{Y}}_{t}   ]  -  \E[ X_k | \X \in  \vec{\mathbb{R}}_n  ] \right |^2 \right ]. \label{eq:boundsOntheDiffNoisyRegime}
\end{align} 
\hrule
\end{figure*}

\section{Proof of Lemma~\ref{lem:EquivalentHighNoise}}
\label{app:EstHighNoise}
From~\eqref{eq:highNoiseRegime}, we have that
\begin{equation}
\label{eq:hkSum}
h_k( \vec{\mathbf{y}}) = \E[ X_k | \X \in  \vec{\mathbb{R}}_n  ] \sum_{ \pi \in  \mathcal{P} }  \mathbb{P}[ \X \in  \vec{\mathbb{R}}_n | \mathbf{Y}= \mathbf{P}_{\pi}\vec{\bf y}].
\end{equation}
We now focus on the summation term in~\eqref{eq:hkSum} and we obtain
\begin{align*}
&\sum_{ \pi \in  \mathcal{P} }  \mathbb{P}[ \X \in  \vec{\mathbb{R}}_n | \mathbf{Y}=\mathbf{P}_{\pi}\vec{\bf y}]
\\ = & \sum_{ \pi \in  \mathcal{P} } \int_{\vec{\mathbb{R}}_n} f_{\X|\Y} (\mathbf{x}|\mathbf{P}_{\pi}\vec{\bf y}) \ {\rm{d}}\mathbf{x}
%
%
\\ \stackrel{{\rm{(a)}}}{=} & \frac{1}{f_{\Y}(\mathbf{y})} \sum_{ \pi \in  \mathcal{P} }  \int_{\vec{\mathbb{R}}_n} f_{\Y|\X} (\mathbf{P}_{\pi}\vec{\bf y}|\mathbf{x}) f_{\X}(\mathbf{x}) \ {\rm{d}}\mathbf{x}
\\ \stackrel{{\rm{(b)}}}{=} & \frac{1}{f_{\Y}(\mathbf{y})}   \int_{\vec{\mathbb{R}}_n} \left( \sum_{ \pi \in  \mathcal{P} } f_{\Y|\X} (\mathbf{P}_{\pi}\vec{\bf y}|\mathbf{x}) \right) f_{\X}(\mathbf{x}) \ {\rm{d}}\mathbf{x}
%
%
\\ \stackrel{{\rm{(c)}}}{=} & \frac{1}{n! f_{\Y}(\mathbf{y})}   \int_{\vec{\mathbb{R}}_n} f_{\vec{\mathbf{Y}} |  \vec{\mathbf{X}} }( \vec{\mathbf{y}} | {\mathbf{x}}) f_{\vec{\X}}(\mathbf{x}) \ {\rm{d}}\mathbf{x}
\\ \stackrel{{\rm{(d)}}}{=} & \frac{1}{n! f_{\Y}({\mathbf{y}})}   f_{\vec{\Y}}(\vec{\mathbf{y}})
%
%
\\ \stackrel{{\rm{(e)}}}{=} & 1,
\end{align*}
where the labeled equalities follow from:
$\rm{(a)}$ Bayes' theorem and since $\Y$ is exchangeable;
$\rm{(b)}$ swapping the role of the sum and integration;
$\rm{(c)}$ using~\eqref{eq:sortingChannelProof} in Lemma~\ref{lem:forMAP} and~\eqref{eq:exch_dist} in Lemma~\ref{lem:Lemma_exchangeable};
$\rm{(d)}$ using the definition of marginal PDF;
and
$\rm{(e)}$ using~\eqref{eq:exch_dist} in Lemma~\ref{lem:Lemma_exchangeable}
and since $\Y$ is an exchangeable random variable.
This concludes the proof of Lemma~\ref{lem:EquivalentHighNoise}.

\section{Proof of Theorem~\ref{thm:HighSigmaRegimeGeneral}}
\label{app:PerfHighNoiseEst}
By using the Pythagorean theorem, we obtain~\eqref{eq:boundsOntheDiffNoisyRegime} at the top of the next page,
where the labeled (in)-equalities follow from:
$\rm{(a)}$ using the optimal estimator in~\eqref{eq:EstimatorBothXandYsymmetric} and the suboptimal estimator in~\eqref{eq:EquivalentHighNoise};
$\rm{(b)}$ the fact that $\sum_{ \pi \in  \mathcal{P} }  \mathbb{P}[ \X \in  \vec{\mathbb{R}}_n | \mathbf{Y}=\mathbf{P}_{\pi}\vec{\bf y}] =1$, as proved in Appendix~\ref{app:EstHighNoise};
$\rm{(c)}$ using Cauchy-Schwarz inequality;
and $\rm{(d)}$ the fact that the probability is less than or equal to one.

We now show that~\eqref{eq:boundsOntheDiffNoisyRegime} goes to zero as $t\to \infty$. We have
\begin{align*}
&\lim_{t\to \infty} \!\mathbb{E} \left [ \left |  \E[ X_k | \X \!\in\!  \vec{\mathbb{R}}_n, \mathbf{Y}_t\!=\!\mathbf{P}_{\pi} \vec{\mathbf{Y}}_{t}   ]  \!-\!  \E[ X_k | \X \!\in\!  \vec{\mathbb{R}}_n  ] \right |^2 \right ]\notag\\
\stackrel{{\rm{(a)}}}{=}& \mathbb{E} \! \left [  \lim_{t\to \infty} \left |  \E[ X_k | \X \!\in\!  \vec{\mathbb{R}}_n, \mathbf{Y}_t\!=\!\mathbf{P}_{\pi} \vec{\mathbf{Y}}_{t}   ]  \!-\!  \E[ X_k | \X \!\in\!  \vec{\mathbb{R}}_n  ] \right |^2 \right ]  
\notag \\
%
%
\stackrel{{\rm{(b)}}}{=} & 0,
\end{align*}
where the labeled equalities follow from:
${\rm{(a)}}$ exchanging the expectation and the limit; this follows from the dominated convergence theorem by using the bound
\begin{align*}
 &\left |  \E[ X_k | \X \in  \vec{\mathbb{R}}_n, \mathbf{Y}_t=\mathbf{P}_{\pi} \vec{\mathbf{Y}}_{t}   ]  -  \E[ X_k | \X \in  \vec{\mathbb{R}}_n  ] \right |^2  \notag\\
 \le & 2 \left( \E[ X_k^2 | \X \in  \vec{\mathbb{R}}_n, \mathbf{Y}_t=\mathbf{P}_{\pi} \vec{\mathbf{Y}}_{t}   ]  +  \E[ X_k^2 | \X \in  \vec{\mathbb{R}}_n  ]  \right);
\end{align*} 
and $\rm{(b)}$ from the assumption in~\eqref{eq:ConitinoutyAssumption}. 
This concludes the proof of Theorem~\ref{thm:HighSigmaRegimeGeneral}.

\section{Proof of Theorem~\ref{thm:boundForVar}}
\label{app:OrdStatMomVar}

\subsection{Proof of~\eqref{eq:ExpOrdStat}}

We have
\begin{align}
\E[X_{(i)}] 
&\stackrel{{\rm{(a)}}}{=} \E[\Phi^{-1}(U_{(i)})] \notag\\
&\stackrel{{\rm{(b)}}}{=} \E[  \Phi^{-1}(\E[U_{(i)}])+  R_0(U_{(i)})] \notag\\
& \stackrel{{\rm{(c)}}}{=}   \Phi^{-1}(\E[U_{(i)}])  + \E[  R_0(U_{(i)})] \notag \\
& \stackrel{{\rm{(d)}}}{=}    \Phi^{-1} \left ( \frac{i}{n+1} \right )  + \E[  R_0(U_{(i)})] \label{eq:expOrdX},
\end{align} 
where the equalities follow from: $\rm{(a)}$~\cite[eq.(2.3.7)]{david2004order}
\begin{align*}
\left ( X_{(1)}, \ldots, X_{(n)} \right ) \stackrel{d}{=} \left(\Phi^{-1}(U_{(1)}), \ldots , \Phi^{-1}(U_{(n)})\right ),
\end{align*}
where $\mathbf{U}$ is standard uniform; moreover, $U_{(i)}$ is a beta $\beta(i,n-i+1)$ variate;
$\rm{(b)}$ the fundamental theorem of calculus, i.e., 
\begin{align*}
  \Phi^{-1}(U_{(i)})-\Phi^{-1}(\E[U_{(i)}])= \underbrace{\int_{\E[U_{(i)}]}^{U_{(i)}}   \frac{ {\rm d} }{{\rm d} x} \Phi^{-1}(x) \ {\rm d} x}_{R_0(U_{(i)})};
\end{align*}
$\rm{(c)}$ linearity of expected value;
$\rm{(d)}$ $U_{(i)}$ is a beta $\beta(i,n-i+1)$ variate, whose expected value is $\E[U_{(i)}]=\frac{i}{n+1}$.

Thus, from~\eqref{eq:expOrdX} we obtain
\begin{align*}
\left | \E[X_{(i)}]  -  \Phi^{-1} \left ( \frac{i}{n+1} \right ) \right |  = \left |\E[  R_0(U_{(i)})] \right |.
\end{align*}
Therefore, we now focus on upper bounding $\left |\E[  R_0(U_{(i)})] \right |$.
First, let $f(\cdot)$ be the PDF of the standard Gaussian random variable, and note that
\begin{align*}
 R_0(u) &= \int_{\E[U_{(i)}]}^{U_{(i)}}   \frac{{\rm d}}{{\rm d}x} \Phi^{-1}(x) {\rm d} x\\
    &=\! (U_{(i)}\!-\! \E[U_{(i)}]) \! \int_{0}^1  \!\!\frac{1}{ f(\Phi^{-1}( tU_{(i)}+(1\!-\!t)\E[U_{(i)}] ))}  {\rm d} t ,
\end{align*} 
where in the last step, we have used the  property 
\begin{align}
&\frac{{\rm d}}{{\rm d}u}\Phi^{-1}(u)=  \frac{1}{ \frac{{\rm d}}{{\rm d}u}\Phi(\Phi^{-1}(u))} \nonumber
\\ =&  \frac{1}{ f(\Phi^{-1}(u))}  =\sqrt{2 \pi} \eu^{ \frac{ (\Phi^{-1}(u))^2}{2}},
\label{eq:ProfInvFun}
\end{align}
and the change of variable $x = tU_{(i)}+(1-t)\E[U_{(i)}]$. Thus,
\begin{align*}
& \left |\E[  R_0(U_{(i)})] \right | 
\\ =& \left | \E \left [ (U_{(i)}\!-\! \E[U_{(i)}]) \! \int_{0}^1  \!\!\frac{1}{ f(\Phi^{-1}( tU_{(i)}+(1\!-\!t)\E[U_{(i)}] ))}  {\rm d} t \right ]  \right |
\\ \stackrel{{\rm{(e)}}}{\leq} & \! \E \! \left [ \left | (U_{(i)}\!-\! \E[U_{(i)}]) \right | \left| \int_{0}^1  \!\!\!\frac{1}{ f(\Phi^{-1}( tU_{(i)}\!+\!(1\!-\!t)\E[U_{(i)}] ))}  {\rm d} t \right| \! \right ],
\end{align*}
where
$\rm{(e)}$ follows by using Jensen's inequality. We now focus on upper bounding the integral term. We have
\begin{align*}
& \int_{0}^1  \!\!\frac{1}{ f(\Phi^{-1}( tU_{(i)}+(1\!-\!t)\E[U_{(i)}] ))}  {\rm d} t 
\\  \stackrel{{\rm{(f)}}}{\leq} & \sqrt{ \frac{\pi}{2} }  \int_{0}^1  \frac{1}{\min \{  x,  1-x    \}}  {\rm d} t 
\\  \stackrel{{\rm{(g)}}}{\leq} & \sqrt{ \frac{\pi}{2} }  \int_{0}^1  \left( \frac{1}{  x(t)}  + \frac{1}{  1-x(t)} \right)  {\rm d} t 
 \\ = & - \sqrt{ \frac{\pi}{2} } \frac{\log \left (  \frac{(U_{(i)}-1)\E[U_{(i)}]}{U_{(i)}(\E[U_{(i)}]-1)} \right )}{U_{(i)}- \E[U_{(i)}]}
\end{align*}
where the equalities/inequalities follow from:
$\rm{(f)}$ by letting $x(t) = tU_{(i)}+(1-t)\E[U_{(i)}] $,
and since
\begin{equation}
f\left( \Phi^{-1}(u) \right ) \geq \sqrt{\frac{2}{\pi}} \min \{u,1-u\}, \ u \in [0,1], \label{eq:LowerBoundOnInverseCDF}
\end{equation}
which follows from the fact that $f\left( \Phi^{-1}(u) \right ) $ is a concave and symmetric function \cite[Lemma~10.1]{boucheron2013concentration}; 
and $\rm{(g)}$ the fact that
$
\min \{a,b\} \geq \frac{ab}{a+b}.
$
%
Therefore, we obtain
\begin{align*}
&   \sqrt{ \frac{2}{\pi} } \left |\E[  R_0(U_{(i)})] \right |  
\\ \leq &  \E \left [ \left | (U_{(i)}\!-\! \E[U_{(i)}]) \right | \! \left | \frac{\log \left (  \frac{(U_{(i)}-1)\E[U_{(i)}]}{U_{(i)}(\E[U_{(i)}]-1)} \right )}{U_{(i)}- \E[U_{(i)}]} \right | \right ]
\\  \stackrel{{\rm{(h)}}}{\leq} & \sqrt{\E \left[  \left| \log \left(  \frac{1-U_{(i)}}{U_{(i)}} \right) + \log \left(  \frac{\E[U_{(i)}]}{1-\E[U_{(i)}]} \right) \right|^2  \right]}
\end{align*}
where ${\rm{(h)}}$ follows from Jensen's inequality.
Now, observe that 
\begin{align*}
&\E \left[  \left| \log \left(  \frac{1-U_{(i)}}{U_{(i)}} \right) + \log \left(  \frac{\E[U_{(i)}]}{1-\E[U_{(i)}]} \right) \right|^2  \right] \\
=&    \mathsf{Var}  \left( \log \left(  \frac{1-U_{(i)}}{U_{(i)}} \right)  \right)  \notag\\
&+  \left( \E \left[   \log \left(  \frac{1-U_{(i)}}{U_{(i)}} \right) \right]+ \log \left(  \frac{\E[U_{(i)}]}{1-\E[U_{(i)}]} \right)      \right)^2.
\end{align*} 
This follows from the property
$\E \left [ \left | X-c\right |^2\right ] = \mathsf{Var} [X] +\left( \E[X]-c\right )^2,$
where $c$ is a constant.
Moreover, for $U_{(i)}$, i.e., a beta $\beta(i,n-i+1)$ variate, we have~\cite{ferrari2004beta} 
\begin{align*}
\mathsf{Var}  \left( \log \left(  \frac{1-U_{(i)}}{U_{(i)}} \right)  \right)& =  \psi_1(i)+ \psi_1(n-i+1),\\
 \E \left[   \log \left(  \frac{1-U_{(i)}}{U_{(i)}} \right) \right]&= \psi(n-i+1)- \psi(i), \\
  \log \left(  \frac{\E[U_{(i)}]}{1-\E[U_{(i)}]} \right)  &= \log \left(  \frac{i}{n+1-i} \right) ,
\end{align*} 
where $\psi_1(\cdot)$ is the second of the polygamma functions, and $\psi(\cdot)$ is the digamma function.
Thus, with this we obtain
\begin{align*}
&    \sqrt{ \frac{2}{\pi} } \left  |\E[  R_0(U_{(i)})] \right |  
\\ \leq &  \sqrt{\psi_1(i)+ \psi_1(n-i+1) + \left( \psi(n-i+1)- \psi(i) +\ell \right )^2},
\end{align*}
where
$
\ell = \log \left(  \frac{i}{n+1-i} \right).
$
Finally, observe that~\cite{guo2015sharp}
\begin{align*}
&\psi_1(k) \leq \frac{1}{k} +\frac{1}{k^2} \stackrel{k \ge 1 }{\leq}  \frac{2}{k}, \\
&\log(x)- \frac{1}{x} \le \psi(x)\le \log(x)- \frac{1}{2x}, \, x>0. 
\end{align*}
With this, we obtain
\begin{align*}
&   \sqrt{ \frac{2}{\pi} } \left |\E[  R_0(U_{(i)})] \right |   \leq   \sqrt{  \frac{2}{i} + \frac{2}{n+1-i}  +  \left |   \frac{1}{i}- \frac{1}{2(n+1-i)}  \right|^2 }.
\end{align*}
This concludes the proof of~\eqref{eq:ExpOrdStat}.

\subsection{Proof of~\eqref{eq:VarOrdStat}}
We have that 
\begin{align}
& \left| \mathsf{Var}(\vec{\mathbf{X}}) - \left(  n - \sum_{i=1}^n  \left( \Phi^{-1} \left ( \frac{i}{n+1} \right )  \right)^2\right) \right|  \notag\\
 \stackrel{{\rm{(a)}}}{=} & \left | - \sum_{i=1}^n  \E^2[X_{(i)}] + \sum_{i=1}^n  \left( \Phi^{-1} \left ( \frac{i}{n+1} \right )  \right)^2 \right |  \notag \\
%
%
 \stackrel{{\rm{(b)}}}{=}  & \left |\sum_{i=1}^n \left [ e_i \left (e_i+2 \Phi^{-1} \left ( \frac{i}{n+1} \right ) \right ) \right ]\right | \notag \\
%
%
 \le &      2 \sum_{i=1}^n \left| \Phi^{-1} \left ( \frac{i}{n+1} \right ) \right| |e_i|   +  \sum_{i=1}^n e_i^2 \notag \\
 \stackrel{{\rm{(c)}}}{\leq}  & 2  \sqrt{ \sum_{i=1}^n \left ( \Phi^{-1} \left ( \frac{i}{n+1} \right ) \right)^2 } \sqrt{ \sum_{i=1}^n e_i^2} +  \sum_{i=1}^n e_i^2, \label{eq:BoundOnTheBariance1}
\end{align} 
where the labeled (in)-equalities follow from: $\rm{(a)}$ the fact that  $\mathsf{Var}(\vec{\mathbf{X}})=n - \sum_{i=1}^n  \E^2[X_{(i)}]$; 
${\rm{(b)}}$ defining 
\begin{equation*}
e_i= \E[X_{(i)}]  -  \Phi^{-1} \left ( \frac{i}{n+1} \right );
\end{equation*}
and 
$\rm{(c)}$ using Cauchy-Schwarz inequality. 

We now bound each term in \eqref{eq:BoundOnTheBariance1}. Towards this end, we use the following lemma, whose proof is provided in Appendix~\ref{app:UsefBoundEps}.
  \begin{lem}\label{lem:BOundOnTHeSumOfQunatileFunction}  
For any $n$ and any $ \epsilon  \ge 0$\footnote{Note that the case $\epsilon=4$ should be computed by taking the limit.}
  \begin{align*}
   & \sum_{i=1}^n  \left| \Phi^{-1} \left ( \frac{i}{n+1} \right )  \right|^{2\epsilon} 
\\ \le &   2^{\frac{11}{4}\epsilon +1}
(n+1)^{\frac{\epsilon}{4}} \left( 1+  \frac{ \left(  \lceil \frac{n+1}{2} \rceil \right)^{1-\frac{\epsilon}{4}}-1}{1-\frac{\epsilon}{4}} \right).
     \end{align*} 
   \end{lem} 
By using Lemma~\ref{lem:BOundOnTHeSumOfQunatileFunction}, we have the following bound
\begin{align}
&\sum_{i=1}^n \left ( \Phi^{-1} \left ( \frac{i}{n+1} \right ) \right)^2 \notag \\
 \leq & 2^{\frac{15}{4}}
(n+1)^{\frac{1}{4}} \left( 1+  \frac{ \left(  \lceil \frac{n+1}{2} \rceil \right)^{\frac{3}{4}}-1}{\frac{3}{4}} \right) \notag \\
 \!\leq\! & \frac{2^{\frac{23}{4}}}{3} (n+1)^{\frac{1}{4}} \left \lceil \frac{n+1}{2} \right \rceil^{\frac{3}{4}} 
 \!\leq\!  \frac{2^{\frac{23}{4}}}{3} (n+1)  \le 18 (n+1). 
\label{eq:BoundOnCDFSUM}
\end{align} 
Then, by using~\eqref{eq:ExpOrdStat}, we obtain
\begin{align}
\sum_{i=1}^n e_i^2 &\le \frac{\pi}{2} \sum_{i=1}^n  \left(  \frac{2}{i} + \frac{2}{n+1-i}  +  \left |   \frac{1}{i}- \frac{1}{2(n+1-i)}  \right|^2  \right) \notag \\
%
%
%
&\stackrel{{\rm{(d)}}}{\le} \frac{3 \pi}{2} \sum_{i=1}^n  \left(  \frac{1}{i} + \frac{1}{n+1-i} \right) \notag \\
%
%
& \stackrel{{\rm{(e)}}}{\le}   \frac{3\pi}{2} \left( 2 \log(n) +  \frac{1}{n}  +1\right), \label{eq:BOundOnSumOFERROTerms}
\end{align} 
where the labeled inequalities follow from:
$\rm{(d)}$ for $a,b$ such that $\min \{a,b\}>0$, we have $|a-b|^2 =(a-b)^2 \leq a^2 + b^2$;
and $\rm{(e)}$ using the integral approximation lemma, namely
\begin{equation}
\label{eq:IntApp}
\sum_{i=N}^M f(n) \leq f(N)+ \int_N^M f(x) \ {\rm{d}}x,
\end{equation}
where $f(\cdot)$ is a monotone decreasing function.

Now, by collecting~\eqref{eq:BoundOnTheBariance1},~\eqref{eq:BoundOnCDFSUM} and~\eqref{eq:BOundOnSumOFERROTerms} we have that 
\begin{align*}
& \left| \mathsf{Var}(\vec{\mathbf{X}}) - \left(  n - \sum_{i=1}^n  \left( \Phi^{-1} \left ( \frac{i}{n+1} \right )  \right)^2\right) \right| \notag\\
   \le & 2  \sqrt{ 18 (n+1) \frac{3\pi}{2} \left( 2 \log(n) +  \frac{1}{n}  +1\right) } 
\\& \quad +  \frac{3\pi}{2} \left( 2 \log(n) +  \frac{1}{n}  +1\right) .
 \end{align*}
This concludes the proof of~\eqref{eq:VarOrdStat}, and hence of of Theorem~\ref{thm:boundForVar}.

\section{Proof of Lemma~\ref{lem:BOundOnTHeSumOfQunatileFunction}}
\label{app:UsefBoundEps}
 First,  observe that from~\eqref{eq:ProfInvFun}, we have that
 \begin{equation}
  \left| \Phi^{-1} \left ( u \right )  \right|^{2} = 2     \log \left(   \frac{1}{\sqrt{2 \pi}}   \frac{1}{ f\left( \Phi^{-1}(u) \right ) }   \right), u \in (0,1),
\label{eq:ReWringSqqureCDF}
 \end{equation} 
where $f(\cdot)$ is the PDF of the standard Gaussian random variable.
Now, we have that
   \begin{align}
 &\sum_{i=1}^n  \left| \Phi^{-1} \left ( \frac{i}{n+1} \right )  \right|^{2\epsilon}  \notag\\
  \stackrel{{\rm{(a)}}}{\le} & 2 \sum_{i=1}^{ \lceil \frac{n+1}{2} \rceil}   \left| \Phi^{-1} \left ( \frac{i}{n+1} \right )  \right|^{2 \epsilon} \notag \\
  \stackrel{{\rm{(b)}}}{\le} & 2^{\epsilon +1} \sum_{i=1}^{ \lceil \frac{n+1}{2} \rceil}  \left(  \log \left(  \frac{1}{\sqrt{2 \pi}}  \frac{1}{ \sqrt{ \frac{2}{\pi} }  \frac{i}{n+1} }   \right) \right)^{\epsilon} \notag \\
%
%
   \stackrel{{\rm{(c)}}}{\le} &  2^{\frac{11}{4}\epsilon +1} \left( n+1 \right)^{\frac{\epsilon}{4}} \sum_{i=1}^{ \lceil \frac{n+1}{2} \rceil}   \frac{1}{  i^{\frac{\epsilon}{4}} } \notag \\
%
%
%
                 \stackrel{{\rm{(d)}}}{\le}& 2^{\frac{11}{4}\epsilon +1}
(n+1)^{\frac{\epsilon}{4}} \left( 1+  \frac{ \left(  \lceil \frac{n+1}{2} \rceil \right)^{1-\frac{\epsilon}{4}}-1}{1-\frac{\epsilon}{4}} \right) \notag,
  \end{align}
  where the labeled (in)-equalities follow from: $\rm{(a)}$  the symmetry of  $ \left| \Phi^{-1} \left ( \frac{i}{n+1} \right )  \right|$; $\rm{(b)}$  using~\eqref{eq:ReWringSqqureCDF} and~\eqref{eq:LowerBoundOnInverseCDF} by noting that $i \leq \frac{n+1}{2}$;  
${\rm{(c)}}$ using the upper bound 
$\log(x) \leq \frac{x^s}{s}, \ \forall s >0,$
with $s=1/4$;
and $\rm{(d)}$ using the integral approximation in~\eqref{eq:IntApp}.
This concludes the proof of Lemma~\ref{lem:BOundOnTHeSumOfQunatileFunction}.

  \section{Proof of Lemma~\ref{lem:ExchangeOflImit}}
\label{lem:LimExp}
Let $V_n^2= \left |  \Phi^{-1}(U_n)   \right|^2 $.    It is not difficult to check that  $U_n \to U$ in distribution~\cite[Example~8.2.1]{resnick2003probability} and, therefore, by the continuity of $x \to \left |  \Phi^{-1}(x)   \right|^2 $, we have that 
  \begin{equation*}
  \lim_{n \to \infty}V_n^2 =  \left |  \Phi^{-1}(U)   \right|^2,
  \end{equation*}
  in distribution. To conclude the proof it remains to show that 
  \begin{equation}
   \lim_{n \to \infty} \E \left[ \left |  \Phi^{-1}(U_n)   \right|^2    \right]=   \E \left[ \lim_{n \to \infty} \left |  \Phi^{-1}(U_n)   \right|^2    \right]. \label{eq:QuantileLimit}
  \end{equation} 
  This exchange of integration and expectation under convergence in distribution is guaranteed by the Vitaly convergence theorem~\cite[Chapter 6]{folland2013real}, provided that the sequence $\{  V_n^2 \}_{n=1}^\infty$ is uniformly integrable.
  
  To show that $\{  V_n^2 \}_{n=1}^\infty$  is uniformly integrable, we check the crystal ball condition~\cite[Chapter~6.5.1]{resnick2003probability} given by
  \begin{equation}
 \sup_{n \ge 1 } \E \left[V_n^{2+\delta} \right]<\infty,
  \end{equation} 
  for some $\delta>0$.  To see this, we choose $\delta=2$ and we observe that, in view of Lemma~\ref{lem:BOundOnTHeSumOfQunatileFunction} with $\epsilon=2$, we obtain
  \begin{equation*}
  \E \left[V_n^{4} \right] \le \frac{2^{\frac{15}{2}}}{n} (n+1)^{\frac{1}{2}} \left( \left( \left \lceil \frac{n+1}{2} \right \rceil \right )^{\frac{1}{2}}-1 \right),
    \end{equation*} 
  which is clearly bounded.  

Therefore,  $V_n^2$ is uniformly integrable  and the interchange of the expectation and limit  in~\eqref{eq:QuantileLimit} is verified. 
This concludes the proof of Lemma~\ref{lem:ExchangeOflImit}.

\end{appendices}

 \bibliography{refs.bib}
 \bibliographystyle{IEEEtran}

\end{document}